\documentclass[aps,prd,showpacs,nofootinbib,superscriptaddress]{revtex4}
\pdfoutput=1
\usepackage{dcolumn}
\usepackage{amsmath,amssymb,setspace,graphicx,subfigure,epsfig}
\usepackage{amsfonts}
\usepackage{latexsym}
\usepackage{epsfig}
\usepackage{color}
\usepackage{nicefrac}

\newcommand{\dis}[1]{\begin{equation}\begin{split}#1\end{split}\end{equation}}
\newcommand{\be}{\begin{equation}}
\newcommand{\ee}{\end{equation}}
\def\bea{\begin{eqnarray}}
\def\eea{\end{eqnarray}}

\newcommand{\Mp}{M_{\rm p}}

\newcommand{\G}{\Gamma}
\newcommand{\p}{\partial}
\newcommand{\Dn}{\delta N}

\newcommand{\vp}{\varphi}
\newcommand{\vps}{\varphi_*}
\newcommand{\chis}{\chi_*}
\newcommand{\epstar}{\epsilon_\vp^*}
\newcommand{\ep}{\epsilon_\vp}

\newcommand{\ecstar}{\epsilon_\chi^*}
\newcommand{\ec}{\epsilon_\chi}

\newcommand{\etapp}{\eta_{\vp\vp}}
\newcommand{\etapc}{\eta_{\vp\chi}}
\newcommand{\etacc}{\eta_{\chi\chi}}

\newcommand{\fnlf}{ f^{(4)}_{\rm NL} }
\newcommand{\fnl}{f_{\rm NL}}
\newcommand{\fnlfinal}{f^{\rm final}_{\rm NL}}
\newcommand{\nz}{n_\zeta}

\newcommand{\Nchi}{N_\chi}
\newcommand{\Nvp}{N_\vp}
\newcommand{\Nchichi}{N_{\chi\chi}}
\newcommand{\Nvpvp}{N_{\vp\vp}}
\newcommand{\Nchivp}{N_{\vp\chi}}

\def\bkone{{\bf k_1}}
\def\bktwo{{\bf k_2}}

\def\picube{(2\pi)^3}
\newcommand{\sdelta}[1]{\!\delta^{\,3}(\mathbf{#1})}

\newcommand{\calP}{{\cal P}}

\newcommand{\Gvp}{\G_\vp}
\newcommand{\Gchi}{\G_\chi}

\begin{document}
\hfill CERN-PH-TH/2012-168
\title{Reheating, Multifield Inflation and the Fate of the Primordial Observables}

\author{Godfrey Leung}
\email{ppxgl@nottingham.ac.uk}
\affiliation{School of Physics and Astronomy, University of Nottingham, University Park, Nottingham, NG7 2RD, UK}

\author{Ewan R. M. Tarrant}
\email{ppxet@nottingham.ac.uk}
\affiliation{School of Physics and Astronomy, University of Nottingham, University Park, Nottingham, NG7 2RD, UK}

\author{Christian T. Byrnes}
\email{cbyrnes@cern.ch}
\affiliation{CERN, PH-TH Division, CH-1211, Gen\`eve 23, Switzerland}

\author{Edmund J. Copeland}
\email{ed.copeland@nottingham.ac.uk}
\affiliation{School of Physics and Astronomy, University of Nottingham, University Park, Nottingham, NG7 2RD, UK}

\pacs{98.80.Cq }

\date{\today}

\begin{abstract}
We study the effects of perturbative reheating on the evolution of the curvature perturbation $\zeta$, in two--field inflation models. 
We use numerical methods to explore the sensitivity of $\fnl$, $\nz$ and $r$ to the reheating process, and present simple qualitative arguments to explain our results.
In general, if a large non--Gaussian signal exists at the start of reheating, it will remain non--zero at the end of reheating.
Unless all isocurvature modes have completely decayed before the start of reheating, we find that the non--linearity parameter, $\fnl$, can be sensitive to the reheating timescale, and that this dependence is most appreciable for `runaway' inflationary potentials that only have a minimum in one direction. 
For potentials with a minimum in both directions, $\fnl$ can also be sensitive to reheating if a mild hierarchy exists between the decay rates of each field.
Within the class of models studied, we find that the spectral index $\nz$, is fairly insensitive to large changes in the field decay rates, indicating that $\nz$ is a more robust inflationary observable, unlike the non--linearity parameter $\fnl$.
Our results imply that the statistics of $\zeta$, especially $\fnl$, can only be reliably used to discriminate between models of two--field inflation if the physics of reheating are properly accounted for. \\
\textit{Keywords: Perturbative Reheating, Multifield Inflation, Primordial Observables}
\end{abstract}
\maketitle

\section{Introduction}\label{sec:intro}

Inflation has become the dominant paradigm for explaining the generation of the primordial density perturbation $\zeta$, that seeded structure formation, and the Cosmic Microwave Background (CMB) anisotropies. According to the standard inflationary scenario, the universe underwent an early period of superluminal expansion, stretching the primordial density perturbations that were generated by vacuum fluctuations of one or more light scalar fields, beyond the causal horizon. On each scale, these fluctuations were promoted to classical perturbations around the time of horizon exit. Over time they were gravitationally amplified, and eventually re--entered the horizon laying the foundations of all cosmic structure that we observe in the universe today.

Extending the initial work of Guth~\cite{Guth1981Inflationary}, the simplest inflationary mechanism invokes a single scalar field whose associated potential has a region which is sufficiently flat to sustain at least $~60$ $e$--folds of accelerated expansion~\cite{Linde1982New,Albrecht1982Cosmology,Hawking1982Supercooled}, required to solve the horizon, flatness, and relic problems (see, e.g.,~\cite{Guth1981Inflationary,Linde1982New,Lyth2009Primordial}). Whilst single--field slow--roll inflation models are consistent with current observational data, there are many reasons to believe that inflation could have been driven by more than one scalar field: theories beyond the standard model of particle physics such as string theory, supergravity and supersymmetry, generically contain multiple scalar fields. Furthermore, with the possibility of greatly enriched field dynamics, multi--field models can give predictions for key physical observables that may be quite different from single field inflation models, and thus offer the chance of being constrained.

Over the last decade, non--Gaussianity has emerged as a powerful probe that may be used to discriminate between different models of inflation. Once the power spectrum of $\zeta$ is known, the assumption that the perturbations are Gaussian makes it possible to specify all the properties of the distribution. Any information contained in the departure from a perfect Gaussian, non–-Gaussianity, is encoded in higher--order correlation functions. Any detection of primordial non--Gaussianity, quantified using the non--linearity parameter $\fnl$, would rule out the simplest models of single field inflation. 

A plethora of different mechanisms for generating a large $\fnl$ have been proposed in the literature. If the inflaton field has canonical kinetic terms then its perturbations are almost exactly Gaussian at Hubble exit and so any significant non--Gaussianity must be generated on super--Hubble scales~\cite{Seery2005Primordial,Maldacena2003NonGaussian}. Features in the inflaton potential~\cite{Chen2007Large}, the curvaton scenario~\cite{Chambers2010NonGaussianity,Mollerach1990Isocurvature,Lyth2002Generating,Linde2006Curvaton,Malik2006Numerical}, modulated reheating/preheating~\cite{Chambers2008Lattice,Chambers2008NonGaussianity,Enqvist2005NonGaussianity,Jokinen2006Very,Kofman2003Probing,Dvali2004New,Suyama2008NonGaussianity,Byrnes2009Constraints}, and an inhomogeneous end of inflation~\cite{Bernardeau2002NonGaussianity,Lyth2005Generating} all generate a large non--Gaussian signal. It is also possible to generate significant non--Gaussianity during multi--field inflation~\cite{Byrnes2009Large,Byrnes2008Conditions}, for a review, see~\cite{Byrnes2010Review}. 

Regardless of the inflationary model, or how many scalar fields were present during inflation, the universe must eventually evolve to the hot radiation dominated era of the standard Big Bang model. By the time inflation has ended the universe is typically in a highly non--thermal state\footnote{An exception to this is warm inflation~\cite{Moss1985Primordial,Berera1995Warm,Berera1999First}, where relativistic particles are continually produced during inflation.}: the superluminal expansion required to homogenise the universe effectively leaves the cosmos at zero temperature, and so a consistent theory of inflation must also explain how the cosmos was \textit{reheated}. This process, which involves a transfer of energy from the inflating field(s) to the standard model particles, can be very complex and may proceed via a number of different mechanisms depending on the inflationary theory.

One of the dawning realisations over the last two decades has been that the process by which the universe is reheated can have a major impact on physical observables, such as the non--linear parameter $\fnl$, the spectral index $\nz$ and the tensor--to--scalar ratio $r$, that are predicted by the preceding inflationary phase~\cite{Bassett1999General,Bassett1999Metric,Bassett2000Massless,Nambu1997Evolution}. Indeed, a number of recent authors~\cite{Byrnes2009Large,Cicoli2012Modulated,Fonseca2012Primordial,Meyers2011NonGaussianities,Meyers2011Adiabaticity,elliston:2011,Tzavara2011Bispectra,Dias2012Multifield} have cautioned that the physics of any subsequent reheating phase may affect the observational predictions of their inflationary models. Generically, we should expect the inflaton to couple to other fields which do not play any role in driving inflation, and such interactions are unavoidable from an effective field theory perspective. For example, it has been shown that the inclusion of such interactions can lead to particle production effects, which radically modify the phenomenology of some inflation models~\cite{Barnaby2009Particle,Barnaby2009Cosmological,Barnaby2011Large,Battefeld2011Beauty}.

As emphasised in~\cite{elliston:2011}, to connect the physics of inflation with observation, the statistics of $\zeta$ should ideally be followed all the way up until the time of last scattering, where the microwave background anisotropy was imprinted. Without a fundamental UV complete theory describing all early universe physics, this is unfortunately impossible. Thankfully however, we may rely on the fact that in the absence of isocurvature (entropy) modes, the curvature perturbation becomes a conserved quantity on superhorizon scales. This was demonstrated to all orders in cosmological perturbation theory~\cite{Lyth2005General} and was also verified using a gradient expansion method~\cite{Langlois2005Evolution,Rigopoulos2003Separate}. Hence, the statistics of $\zeta$ evaluated when all isocurvature modes are exhausted and an adiabatic condition reached, are those that are measured today. 

Many previous works have assumed that the universe is reheated instantaneously~\cite{Dimopoulos2012Inflating,Choi2012Primordial,Sasaki2008Multibrid,Naruko2009Large}, fossilising the curvature perturbation immediately. But this is an idealisation: reheating presumably takes a finite time to complete. Recently, the authors of~\cite{Hazra2012Scalar} have shown explicitly that for canonical single field inflation models with quadratic minima, an epoch of preheating does not alter the amplitude of the scalar bi--spectrum generated during inflation. Ideally, the evolution of $\zeta$ should also be followed through the subsequent phase of reheating where the energy of the oscillating inflaton is transferred to radiation. However recent studies have shown that the amplitude of the curvature perturbation remains unaffected even during perturbative reheating, see for example~\cite{Jain:2009ep}. Thus naively, one might expect the scalar bi--spectrum to also remain unchanged by this process since $\zeta$ itself is conserved at a non--linear level for single field models, although this remains to be seen explicitly. 

Even less clear is how reheating affects the evolution of $\zeta$ in multi--field models, since when more than one field is present, isocurvature fluctuations can cause $\zeta$ to evolve on super--Hubble scales. It is already known that the two--point correlation function of $\zeta$ can be affected by metric preheating~\cite{Finelli2000Parametric}. Until an adiabatic condition is reached, such as in the case when the universe is radiation dominated, all observable quantities associated with $\zeta$ continue to evolve. How sensitive then are the key inflationary observables to the reheating process? Is this sensitivity heavily dependent on the inflationary model? Does the level of non--gaussianity that exists at the end of inflation survive until the completion of reheating? Focusing on two--field inflation models and assuming that reheating proceeds perturbatively, it is the purpose of this paper to address these questions.

By numerically implementing the $\delta N$ formalism, we follow the evolution of $\zeta$ beyond the end of inflation, until the completion of a phase of perturbative reheating. We parametrise the decay of the oscillating inflation and isocurvature fields into relativistic particles by introducing decay terms into the field equations. Our goal is to investigate the sensitivity of the key inflationary observables to the physics of perturbative reheating. We study two classes of potential: the `runaway' type which has a minimum in only one direction; and potentials which have a minimum in both directions.

The paper is organised as follows: in Section~(\ref{sec:backgroundTheory}) we recall the $\delta N$ formalism and review the textbook elementary theory of reheating, before discussing its numerical implementation within the separate universe picture.  In Section~(\ref{sec:onemin}) we study the evolution of $\fnl$ and other $\zeta$-related statistics during the reheating phase for the class of potentials which posses a minimum in only one direction. Then, in Section~(\ref{sec:twomins}) we repeat the same analysis for potentials where both directions have a minimum. We also consider an example of non--separable potential models in Section~(\ref{sec:non_separable}). We discuss and conclude in Section~(\ref{sec:discConclu}). The expert reader familiar with the elementary theory of reheating and the $\delta N$ formalism may wish to omit Sections~(\ref{sec:reheating}) and~(\ref{sec:deltaN}). 

\section{Perturbative Reheating, Non--Gaussianity and the $\delta$N Formalism}\label{sec:backgroundTheory}

The two--field inflation models that we study in this paper are described by the action
\dis{ 
\label{eq:action}
S=\int d^4 x \sqrt{-g}\left[\Mp^2\frac{R}{2}-\frac12g^{\mu\nu}\partial_\mu\vp\partial_\nu\vp-\frac12g^{\mu\nu}\partial_{\mu}\chi\partial_\nu\chi-W(\vp,\chi) \right]\,,
}
where $\Mp=1/\sqrt{8\pi G}$ is the reduced Planck mass. The standard slow--roll parameters are defined as
\bea
\label{eq:slowrollpar}
\ep = \frac{\Mp^2}{2}\left(\frac{W_{\vp}}{W}\right)^{2}\,, \quad \ec = \frac{\Mp^2}{2}\left(\frac{W_{\chi}}{W}\right)^{2}\,, \quad \epsilon = \ep + \ec \,,\nonumber \\
\etapp=\Mp^2\frac{W_{\vp\vp}}{W}\,, \quad \etapc=\Mp^2\frac{W_{\vp\chi}}{W}\,, \quad \etacc=\Mp^2\frac{W_{\chi\chi}}{W}\,,
\eea
where the subscripts denote differentiation with respect to the fields. 

We will consider various forms of $W(\vp,\chi)$, with the only constraint that $W(\vp,\chi)$ must have a minimum in one, or both of the field directions to enable at least one field to oscillate and reheat the universe. The fundamental difference in form between one--minimum and two--minima potentials provides a logical division of our analysis into classes. This is partly motivated by the work of~\cite{elliston:2011} where two broad classes of behaviour for the evolution of $\zeta$ were recognised: Potentials that contain a `natural focussing region', which is guaranteed for a two--field model with minima in both directions, allow neighbouring trajectories in field space to converge `naturally', quenching the flow of power from isocurvature modes to $\zeta$. Alternatively no such focussing region may exist, which is the case for a two--field model with only a single minimum, and so $\zeta$ will continue to evolve until an adiabatic condition is reached. In the latter case, predictions for observables such as $\fnl$ cannot currently be linked directly to the physics of the inflationary model, as they will be dependent on the subsequent phase of reheating. Even in the former case, if the universe \textit{approaches} adiabaticity by the inflating/isocurvature trajectories converging in, and oscillating about, their global minima, then it is not clear how the decay of the oscillating fields into radiation affects the final stages of the evolution of $\zeta$. We note that adiabaticity may also be reached via a third waterfall field, as is the case in hybrid inflation~\cite{Linde1994Hybrid}.

In the following subsections we introduce the simple perturbative reheating scheme that we use throughout this paper, and briefly review the $\delta$N formalism that is used to compute the statistics of $\zeta$.

\subsection{Elementary theory of reheating}\label{sec:reheating}

In this section we recall the elementary theory of reheating based on perturbation theory that was developed in~\cite{Dolgov1982Baryon,Abbott1982Particle}. For the purposes of this discussion, we assume for simplicity that the potential $W(\vp,\chi)$ in the action Eq.~(\ref{eq:action}) only has a single minimum in, say, the $\chi$ direction, and we assume that this is the inflationary direction. The reheating mechanism presented here only applies to the directions in the potential that are associated with well--defined minima.  At inflationary energy scales we may neglect the contribution to gravity from any other fields such as $\chi_b$ bosons (not to be confused with the inflationary $\chi$ field) or $\psi_f$ fermions in the action Eq.~(\ref{eq:action}). Hence, for cosmological applications we may retain only the dominant fields $\vp$, $\chi$ and gravity. Then in a flat FRW universe, the Friedmann equation reads:
\be 
\label{eq:Hubble1}
H^2=\frac{1}{3\Mp^2}\left[\frac12\dot{\vp}^2+\frac12\dot{\chi}^2+W(\vp,\chi)\right]\,.
\ee
The dynamics of $\chi$ is governed by the Klein--Gordon equation
\be 
\label{eq:KG1}
\ddot{\chi}+3H\dot{\chi}+W_{,\chi}=0\,,
\ee
and similarly for $\vp$. For sufficiently large initial values of $\chi\,,\vp>\Mp$, Hubble friction dominates over $\ddot{\chi}$ (and $\ddot{\vp}$) and the potential term $W(\vp,\chi)$ in Eq.~(\ref{eq:Hubble1}) is assumed to dominate over the kinetic terms. During this slow--roll stage, the universe inflates, expanding quasi--exponentially. As the inflating $\chi$ field rolls toward its minimum at ${\chi}_0$ it gains kinetic energy, eventually bringing inflation to an end, whilst the $\vp$ field continues to contribute to the expansion rate. We assume that the minimum in the $\chi$ direction is quadratic to leading order, $\frac12m^2_\chi\chi^2$, and so ${\chi}_0=0$. We note that a similar discussion may also be applied for theories with quartic minima~\cite{Kofman1996Origin}.  Ignoring for the moment the effects of particle production, as the inflaton approaches and inevitably overshoots its minimum, it begins to oscillate about ${\chi}_0$ on a shorter time scale compared to the Hubble time. Here, we assume that $H\ll m_\chi$ after inflation has ended.\footnote{For potentials with local curvature much different to $\frac12m_\chi^2\chi^2$, this estimate can be very different.} The frequency of the oscillations is $k_0=m_\chi$. The large vacuum energy of the inflaton then exists in spatially coherent oscillations, which can be interpreted as a collection of a number of $\chi$--particles with zero momenta. The density, $n_{\chi}=\rho_\chi/m_\chi$ of this coherent wave of particles decreases as $a^{-3}$, since the condensate behaves as non--relativistic pressureless matter: $\rho_\chi=\frac12(\dot{\chi}^2+m^2_\chi\chi^2)\sim a^{-3}$.

The amplitude of the $\chi$ oscillations gradually decays due to the Hubble expansion and also because of the transfer of energy to lighter particles produced by the oscillating field.  As these decay products thermalise, the Universe is reheated. The inflaton may decay into bosons $\chi_b$ and fermions $\psi_f$ due to $-\frac12g^2\chi^2\chi_b^2$ and $-h\bar{\psi_f}\psi_f\chi$ interaction terms, which should now be included into the fundamental action Eq.~(\ref{eq:action}). Based on the above interpretation of the spatially homogeneous, coherently oscillating $\chi$ field, the effects of particle production may be incorporated into Eq.~(\ref{eq:KG1}) \cite{Linde2005Particle}:
\be 
\label{eq:KGquantum}
\ddot{\chi}+3H\dot{\chi}+\left(m_\chi^2 + \Pi(k_0)\right)\chi=0\,.
\ee
Here, $\Pi(k_0)$ is the flat space polarisation operator for the field $\chi$ at four--momentum $k=(k_0,0,0,0)$. It can be shown that the real part of $\Pi(k_0)$ gives only a small correction to $m_\chi^2$, but when $k_0\ge min(2m_{\chi_b},2m_{\psi_f})$, $\Pi(k_0)$ acquires an imaginary part ${\rm Im}\,\Pi$. Working in the limit $m_\chi\gg \{H,{\rm Im}\,\Pi\}$, which are conditions that should be satisfied after inflation, neglecting the time--dependence of ${\rm Im}\,\Pi$ and assuming $H=2/3t$, the approximate solution to Eq.~(\ref{eq:KGquantum}) is:
\be 
\label{eq:KGquantumSol}
\chi(t)\approx\frac{\Mp}{\sqrt{3\pi}m_\chi t}{\rm exp}\left(-\frac12\G t \right){\rm sin}\,(m_\chi t)\,,
\ee
where $\G=\G(\chi\rightarrow\chi_b\chi_b)+\G(\chi\rightarrow\psi_f\psi_f)$ is the \textit{total} decay rate of $\chi$ particles. Here we have used the relation ${\rm Im}\,\Pi=m_\chi\G$ which follows from unitarity~\cite{pesk}. Eq.~(\ref{eq:KGquantumSol}) implies that the amplitude of the $\chi$ oscillations decays as $\chi(t)\sim a^{-3/2}{\rm exp}(-\frac12\G t)$.\\

For a phenomenological description of the reheating effect, one can add an extra friction term $\Gchi\dot{\chi}$ to the classical equation of motion of the field $\chi$, instead of adding the polarization operator~\cite{Kolb1990Early,Kofman1996Origin}:
\be 
\label{eq:KGpheno}
\ddot{\chi}+(3H+\Gchi)\dot{\chi}+W_{,\chi} =0\,.
\ee
Once again assuming $H=2/3t$ and a quadratic minimum, $W=\frac12m^2_\chi\chi^2$, the solution of this equation is exactly Eq.~(\ref{eq:KGquantumSol}). Multiplying through by $\dot{\chi}$ it is intuitive to rewrite Eq.~(\ref{eq:KGpheno}) as $\dot{\rho}_\chi+3H\dot{\chi}^2+\Gchi\dot{\chi}^2=0$. Now, since $\chi$ is rapidly oscillating around $\chi_0$ approximately sinusoidally, it can be replaced by its average over a single oscillation cycle\footnote{If the motion of $\chi$ is approximately that of a simple harmonic oscillator, $\langle V\rangle=\langle\dot{\chi}^2/2\rangle=\rho_\chi/2$ and so we see that $\langle P_\chi\rangle=\langle\dot{\chi}^2/2-V\rangle$ vanishes and the coherent oscillating $\chi$ behaves as pressureless matter, justifying our previous statements.}, $\langle\dot{\chi}^2\rangle_{\rm cycle}=\rho_\chi$. If the decay products of the oscillating $\chi$ field are very light relative to $\chi$ itself, and are only bosonic, we can model them as a (single) relativistic radiation fluid:
\bea
\label{eq:radiation}
\dot{\rho}_\gamma+4H\rho_\gamma&=&\Gchi\rho_\chi=\Gchi\dot{\chi}^2\,, \\
\label{eq:Hubble2}
H^2&=&\frac{1}{3\Mp^2}\left(\rho_\chi+\rho_\vp+\rho_\gamma\right)\,.
\eea
At this point a number of comments surrounding the validity of Eqs.~(\ref{eq:KGpheno}) and~(\ref{eq:radiation}) are in order. Firstly and most importantly, this simple phenomenological equation~(\ref{eq:KGpheno}) is only valid when $\chi$ is rapidly oscillating about $\chi_0$: the `particle creation' term, $\Gchi\dot{\chi}$, should not be present during inflation. Furthermore, since in this example the $\vp$ field does not have a minimum about which it can oscillate, it should not be coupled to radiation: $\Gvp=0$. Secondly, Eq.~(\ref{eq:KGpheno}) (as is Eq.~(\ref{eq:KGquantum})) is valid only when $m_\chi\gg H$ and $m_\chi\gg\Gchi$. We have also made the assumption that the decay rate $\Gchi$ of the inflaton can be calculated using the standard methods of quantum field theory, describing the decay $\chi\rightarrow\chi_b\chi_b$. If however, many $\chi_b$--particles were produced in the early stages of particle production, the probability of decay becomes greatly enhanced by effects related to Bose--statistics, which may lead to explosive particle production~\cite{Traschen:1990sw,Dolgov:1989us,Kofman:1997yn}. When the amplitude of the oscillating field is sufficiently large, we should also expect reheating to occur in a different way through parametric or stochastic resonance~\cite{Kofman:1997yn,Shtanov1995Universe,Bassett2006Inflation}.

In this perturbative scheme, reheating completes at time $t_c$, when the Hubble rate $H^2=\rho/3\Mp^2\sim t_c^{-2}$ drops below the decay rate $\Gchi$. The density of the universe at this moment is then
\be 
\label{eq:densitiesReheatComplete}
\rho(t_c)\simeq3H^2(t_c)\Mp^2=3\Gchi^2\Mp^2\,.
\ee
If the decay products interact with each other strongly enough, then thermal equilibrium is quickly established and may be maintained at a temperature $T_R$. Treating this ultrarelativistic gas of particles with Bose--Einstein statistics, the energy density of the universe in thermal equilibrium is then
\be 
\label{eq:boseEinstein}
\rho(T_R)\simeq\left(\frac{\pi^2}{30}\right)g_* T_R^4\,,
\ee
where the factor $g_*(T_R)\sim10^2 -10^3$ depends on the number of ultrarelativistic degrees of freedom. Comparing Eqs.~(\ref{eq:densitiesReheatComplete}) and (\ref{eq:boseEinstein}) we arrive at
\be 
\label{eq:reheatTemp}
T_R\sim0.1\sqrt{\Gchi\Mp}\,.
\ee
In order not to spoil the success of BBN, the inflaton decay products should be quickly thermalized through scatterings, annihilations, pair creation and further decays, such that the universe is completely radiation dominated before the BBN epoch. This constrains the reheating temperature to be $T_R\gtrsim5\,$MeV~\cite{Kawasaki1999Cosmological,Ichikawa2005Oscillation}, which in turn implies $\Gchi\gtrsim 4\times10^{-40}\Mp$. We ensure that this bound is always satisfied throughout our paper. For such weak decay rates, reheating would proceed incredibly slowly if the process were entirely perturbative. In reality however, as alluded to above, the universe is unlikely to be reheated via a mechanism that can be described completely by standard perturbation theory, and so we interpret such bounds on $\Gchi$ rather loosely. There is also an upper bound on $T_R$ (and so $\Gchi$) coming from the overproduction of gravitinos~\cite{Kohri2006BigBang,Steffen2008Probing}, which does not apply to us as we are not considering supersymmetric models.

Despite various limitations, the elementary theory of reheating is appealing due to its simplicity and ability to be very successful in describing the reheating process in certain regimes. In this paper we are interested in the effects that reheating has on the evolution of statistical properties of $\zeta$, such as $\fnl$. To this end, we parametrise the reheating process with Eqs.~(\ref{eq:KGpheno}),~(\ref{eq:radiation}) and~(\ref{eq:Hubble2}) and assume that any important physics that may affect the evolution of $\zeta$ are well described by this parametrisation. Since we are concerned with two--field models of inflation, we also assume that this description of reheating applies to both fields, $\vp$ and $\chi$, subject to the limitations discussed above. 

Whilst reheating may well be more complex than the simple perturbative model we consider, it is a useful scheme for determining how sensitive the primordial observables may be to reheating, and to check whether any general trend exists across different models. For example, one might speculate that any large non--Gaussianity is generically damped to zero by reheating, as is often (but not always~\cite{kim2010}) the case during inflation if the isocurvature mode decays during slow--roll~\cite{Meyers2011NonGaussianities,Watanabe2012Delta}. We will show that this is not the case for reheating.

\subsection{The $\delta$N Formalism and non--Gaussianity}\label{sec:deltaN}

The $\delta$N formalism~\cite{Sasaki1996General,Sasaki1998SuperHorizon,Lyth2005Inflationary} has been used extensively throughout the literature to compute the primordial curvature perturbation and its statistics. The formalism relates $\zeta$ to the number of $e$--folds of expansion $N$, given by:
\be 
\label{eq:numEfoldsDef}
N({t_*}\,, {t_c})=\int^{t_c}_{t_*} H(t){\rm d}t,
\ee
which is evaluated from an initial flat hypersurface to a final uniform density hypersurface. The perturbation in the number of $e$--foldings, $\delta N$, is the difference between the curvature perturbations on the initial and final hypersurfaces. We take the initial time to be Hubble exit during inflation, denoted by $t_*$, and the final time, denoted by $t_c$, to be a time deep in the radiation dominated era when reheating has completed. The curvature perturbation is then given by \cite{Lyth2005Inflationary} (or~\cite{Saffin2012Covariance} for the covariant approach)
\be
\label{eq:deltaN} 
\zeta=\delta N= \sum_I
N_{,I}\delta\vp_{I*}+\frac12\sum_{IJ}N_{,IJ}\delta\vp_{I*}\delta\vp_{J*}+\cdots\,, 
\ee
where $N,_I=\partial N/(\partial \vp^I_*)$ and the index $I$ runs over all of the fields.  In general, $N(t_c,t_*)$ depends on the fields, $\vp_I(t)$, and their time derivatives, $\dot{\vp}_I(t)$. However, if the slow--roll conditions, $3H\dot{\vp}_I \simeq - W_{,I}$, are satisfied at Hubble exit, then $N$ depends only on the initial field values. The radiation fluid remains effectively unperturbed at horizon exit as it does not yet exist, and so does not feature in the above expansion. The power spectrum and bispectrum defined (in Fourier space) are given by~\cite{Lyth2009Primordial}:
\begin{eqnarray}
\label{eq:powerspectrumdefn} 
\langle\zeta_{\bkone}\zeta_{\bktwo}\rangle &\equiv&
\picube\,
\sdelta{\bkone+\bktwo}\frac{2\pi^2}{k_1^3}\calP_{\zeta}(k_1) \, , \\
\langle\zeta_{{\mathbf k_1}}\,\zeta_{{\mathbf k_2}}\,
\zeta_{{\mathbf k_3}}\rangle &\equiv& \picube\, \sdelta{{\mathbf
k_1}+{\mathbf k_2}+{\mathbf k_3}} B_\zeta( k_1,k_2,k_3) \,. 
\end{eqnarray}
From this we can define three quantities of key observational interest, respectively the spectral index, the tensor--to--scalar ratio and the non-linearity parameter
\bea 
\label{eq:nzeta}
n_{\zeta}-1&\equiv& \frac{\partial \log\calP_{\zeta}}{\partial\log k}\,, \\
\label{eq:scaltensR}
r&=&\frac{\calP_T}{\calP_{\zeta}}=\frac{8\calP_*}{\Mp^2\calP_{\zeta}}\,, \\
\label{eq:fnldefn} 
\fnl&=&\frac56\frac{k_1^3k_2^3k_3^3}{k_1^3+k_2^3+k_3^3}
\frac{B_{\zeta}(k_1,k_2,k_3)}{4\pi^4\calP_{\zeta}^2}\,. 
\eea
Here $\calP_*$ is the power spectrum of the scalar field fluctuations and $\calP_T=8\calP_*=8H_*^2/(4\pi^2\Mp^2)$ is the power spectrum of the tensor fluctuations. As defined above, $\fnl$ is shape dependent, but it has been shown that the shape dependent part is much less than unity~\cite{Vernizzi2006NonGaussianities,Wands2002Observational,Maldacena2003NonGaussian,Seery2005Primordial,Arroja2008Nongaussianity} for local non--Gaussianity in canonical models. Since the ideal CMB experiment is only expected to reach a precision of $\fnl$ around unity~\cite{Komatsu2001Acoustic}, we calculate the shape independent part of $\fnl$, denoted by $\fnlf$ in \cite{Vernizzi2006NonGaussianities,Choi2007Spectral}. Whenever the non--Gaussianity is large, $|\fnl|>1$, as is the case considered throughout this paper, we can associate $\fnlf\simeq\fnl$. This $k$ independent part of $\fnl$ and the spectral index can be calculated by the $\delta N$ formalism,
\bea
{\cal P}_\zeta&=&\sum_I N_{,I}^2 {\cal P}_*\,,\label{eq:spectrum} \\
n_{\zeta}-1&=& -2\epsilon^* + \frac{2}{H_*}\frac{\sum_{IJ}\dot{\vp}_*{_J} N_{,JI}N_{,I}}{\sum_K N_{,K}^2}\,,\label{eq:index} \\
\fnl&=&\frac56 \frac{\sum_{IJ}N_{,IJ}N_{,I}N_{,J}}{\left(\sum_I N_{,I}^2\right)^2} \label{eq:fnl} \,.
\eea
We use the same sign convention for $\fnl$ as the WMAP team~\cite{Komatsu2011SEVENYEAR}. The latest observations from 7 years of WMAP data are~\cite{Komatsu2011SEVENYEAR}
\bea 
n_{\zeta}&=&0.967^{+0.014}_{-0.014}\qquad (\mathrm{assuming}\,\,r=0)\,, \\
r&<&0.36\,\,\,\,(95\%\,\,\rm{CL})\,, \\ -10&<&\fnl^{\textrm{local}}<74
\,\,\,\,(95\%\,\,\rm{CL})\,. 
\eea

The crucial difference between single and multi--field inflation is that in single field inflation, the slow--roll solution forms a one--dimensional phase space. Hence, by virtue of the attractor theorem there is a unique inflationary trajectory that is always quickly reached. Furthermore, the end of inflation takes place at a fixed value of the inflaton field, corresponding to a fixed energy density. When two fields are present however, the phase--space is two--dimensional with an infinite number of possible classical trajectories in field space. The values of the two fields at the end of inflation will in general depend on the choice of trajectory. Then, to compute the $\delta N$ derivatives ($N_{,I}$ etc) in multi--field models, an extra piece of information, a conserved quantity along a given trajectory, is required. Within slow--roll, such a constant of motion exists (see for example~\cite{Vernizzi2006NonGaussianities,Wang2010Note,Meyers2011NonGaussianities}), and under the assumption that the potential is sum separable $W= \lambda(U(\vp)+V(\chi))^a$, or product separable $W= \lambda(U(\vp)V(\chi))^a$ in the fields, explicit expressions for the $\delta N$ derivatives may be obtained~\cite{Vernizzi2006NonGaussianities,Choi2007Spectral,Byrnes2008Conditions,Meyers2011NonGaussianities}. 

Recently, the authors of~\cite{Seery2012Inflationary} used raytracing techniques to reformulate inflationary perturbation theory in the language of geometrical optics. Whilst this technique yields differential equations from which the $\delta N$ coefficients can be computed efficiently, closed--form expressions for the resultant path--ordered exponential integrals can only be obtained under conditions of separability and slow--roll. By decomposing the field perturbations into curvature and isocurvature perturbations, similar expressions are also found in~\cite{Mazumdar2012Separable}.

Exact solutions, valid beyond slow--roll, have been obtained assuming a sum--separable ansatz for the Hubble parameter~\cite{Byrnes2009NonGaussianity}. Such an ansatz, besides being very restrictive, cannot be applied to a phase of perturbative reheating as it relies upon monotonicity of the field variables. 

Alternative long--wavelength (LWL) formulae have also been developed to analytically study the nonlinear evolution of long wavelength cosmological perturbations in the early universe ~\cite{Kodama:1997qw}. In such an approach, the perturbations are written in terms of quantities of the corresponding exactly homogenous universe to the leading order of the gradient expansion. The formulae have recently been extended to study nonlinear perturbations in universe where multiple scalar fields and perfect fluids coexist~\cite{Hamazaki:2008mh,Hamazaki:2011en}.

In this paper, we will use the $\delta$N formalism. In all cases, what currently evades us is a practical method for analytically computing the $\delta$N derivatives for \textit{arbitrary} potentials during slow--roll and separable (and non--separable) potentials \textit{beyond} slow--roll. While still assuming slow--roll at Hubble exit however, one can go beyond the slow--roll approximation and the condition of separability by numerically solving the second order equations of motion, Eqs.~(\ref{eq:KGpheno}) and~(\ref{eq:radiation}), together with the Friedman constraint~(\ref{eq:Hubble2}), introducing the decay terms $\Gvp$ and $\Gchi$ when applicable.

\subsection{Numerical Code}\label{sec:numericalcode}

The $\delta N$ formalism is based on the assumption that (smoothed) spatially separated patches of the universe will evolve on super--horizon scales like independent, unperturbed universes up to small corrections. This is the separate universe picture~\cite{Bardeen1983Spontaneous,Lyth1985LargeScale}. An ensemble of smoothed regions picks out a collection of trajectories in phase space which is often referred to as a `bundle'~\cite{elliston:2011,Seery2012Inflationary}. In essence, the $\delta N$ formalism requires knowledge about how such a bundle, centred on a fiducial trajectory, evolves. Our choice of gauge demands that each trajectory in the bundle is evolved from an initially flat hypersurface up to a hypersurface of constant energy density. Hence, each trajectory will experience a slightly different expansion history in order to bring them to a common energy density. The adiabatic mode is generated by fluctuations along the fiducial trajectory, whilst fluctuations between neighbouring trajectories generate the isocurvature modes.

Acknowledging this simple picture, the $\delta N$ formalism may be implemented numerically: First, the fiducial trajectory emanating from $\{\vps,\chis\}$ is constructed by solving the full, non--linear second order field equations, i.e., Eq.~(\ref{eq:KGpheno}) (and similarly for the $\vp$ field) together with the Friedmann constraint~(\ref{eq:Hubble2}) and the equation for the radiation fluid~(\ref{eq:radiation}). The bundle is then formed by evolving neighbouring trajectories with slightly perturbed initial conditions, $\vps\rightarrow\vps+\delta\vps$ and $\chis\rightarrow\chis+\delta\chis$. Each trajectory in the bundle is then brought to a common energy hypersurface where the partial derivative of $N(t_c,t_*)$ with respect to the field values at horizon crossing $\{\vps,\chis\}$ is taken using a seven--point `stencil' finite difference method~\cite{Abramowitz1965Handbook}. This provides a fast, efficient method for computing $\nz$, $r$ and $\fnl$ for an arbitrary two--field model, valid beyond slow--roll and through a phase of reheating. Numerical codes based on the moment transport equations have also been developed~\cite{Mulryne2011Moment}.

As discussed in Section~(\ref{sec:reheating}), the reheating parameters $\Gvp$ and $\Gchi$ are set to zero during inflation. It is only when each individual trajectory in the bundle passes through its minimum $\{\chi_0,\vp_0\}$ for the first time that $\Gvp$ and $\Gchi$ are introduced to the field equations, sourcing the radiation fluid. In general, for any given trajectory, $\vp$ will not reach the minimum of its potential at the same time as $\chi$, and so $\Gvp$ and $\Gchi$ are `switched on' at different times along the same trajectory. Furthermore, for each directions in the potential, the foliation of the entire bundle of trajectories as determined by each trajectory reaching $\chi_0$ (and likewise $\vp_0$) does not in general occur at a surface of constant time or a surface of constant energy, but rather at a surface of constant $\chi_0$ (and $\vp_0$)\footnote{This is true for global minima. If the oscillations of one field, $\chi$ say, occurred in a local minimum, which is a function of the other field, $\chi_0(\vp)$, this statement will not hold true. We do not consider such models in this paper.}. We refer to these surfaces as the \textit{reheating hypersurfaces}. For potentials which have minima in both directions there are two such hypersurfaces. If the potential does not have a minimum in the $\chi$ (or $\vp$) direction, then $\Gchi=0$ (or $\Gvp=0$) always. Furthermore, we also ensure that when the potential has a minimum in, say, the $\chi$ direction, the conditions $m_\chi\gg\Gchi$ and $m_\chi\gg H$ are satisfied. This definition of the reheating hypersurface is more refined than that of~\cite{elliston:2011}, where reheating was initiated at a surface of constant density. It is also different to that of~\cite{Mulryne2011NonGaussianity}, where the decay terms were present throughout inflation. The rest of this paper is dedicated to exploring the sensitivity of $\fnl$, $\nz$ and $r$ to the reheating process.

\section{One Minimum}\label{sec:onemin}

In this section we present numerical results for the statistics of $\zeta$ for the class of two--field potentials which have one minimum in the $\chi$ direction. In what follows, $\chi$ may be identified as the inflaton and $\vp$ as the field which sources the isocurvature perturbations. The $\vp$ field is not directly involved in the reheating phase and so $\Gvp=0$ at all times.

\subsection{Quadratic minimum: $W(\vp,\chi)=W_0\chi^2e^{-\lambda\vp^2}$}\label{sec:oneminquadratic}

This potential was first introduced by~\cite{Byrnes2008Conditions}, and has made frequent appearances in the literature since then~\cite{elliston:2011,Dias2012Transport,Huston2012Calculating,Anderson2012Transport,Watanabe2012Delta,Peterson2011NonGaussianity}. It does not contain a `focussing' region where neighbouring trajectories in the bundle may converge, due to its `runaway' form in the $\vp$ direction. Hence, $\zeta$ and its statistics will continue to evolve after inflation has ended. The parameter space for which $\fnl$ may be large at the end of inflation was derived in~\cite{Byrnes2008Conditions}. Essentially, the initial background trajectory must be fined--tuned to be nearly parallel to the axis of the inflaton.
\begin{figure*}[!htb]
	\begin{tabular}{cc}
		\includegraphics[width=0.48\linewidth]{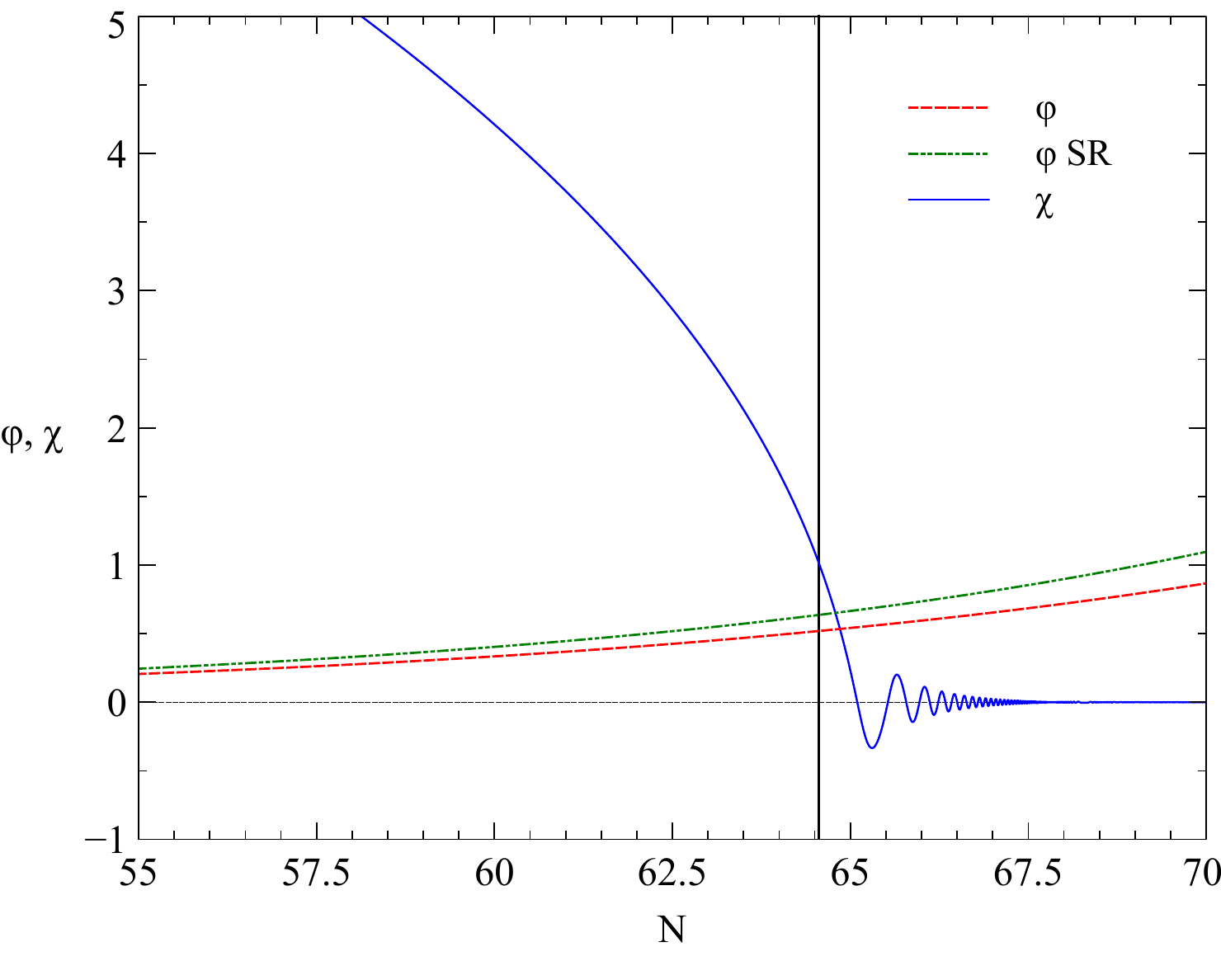}
		\includegraphics[width=0.48\linewidth]{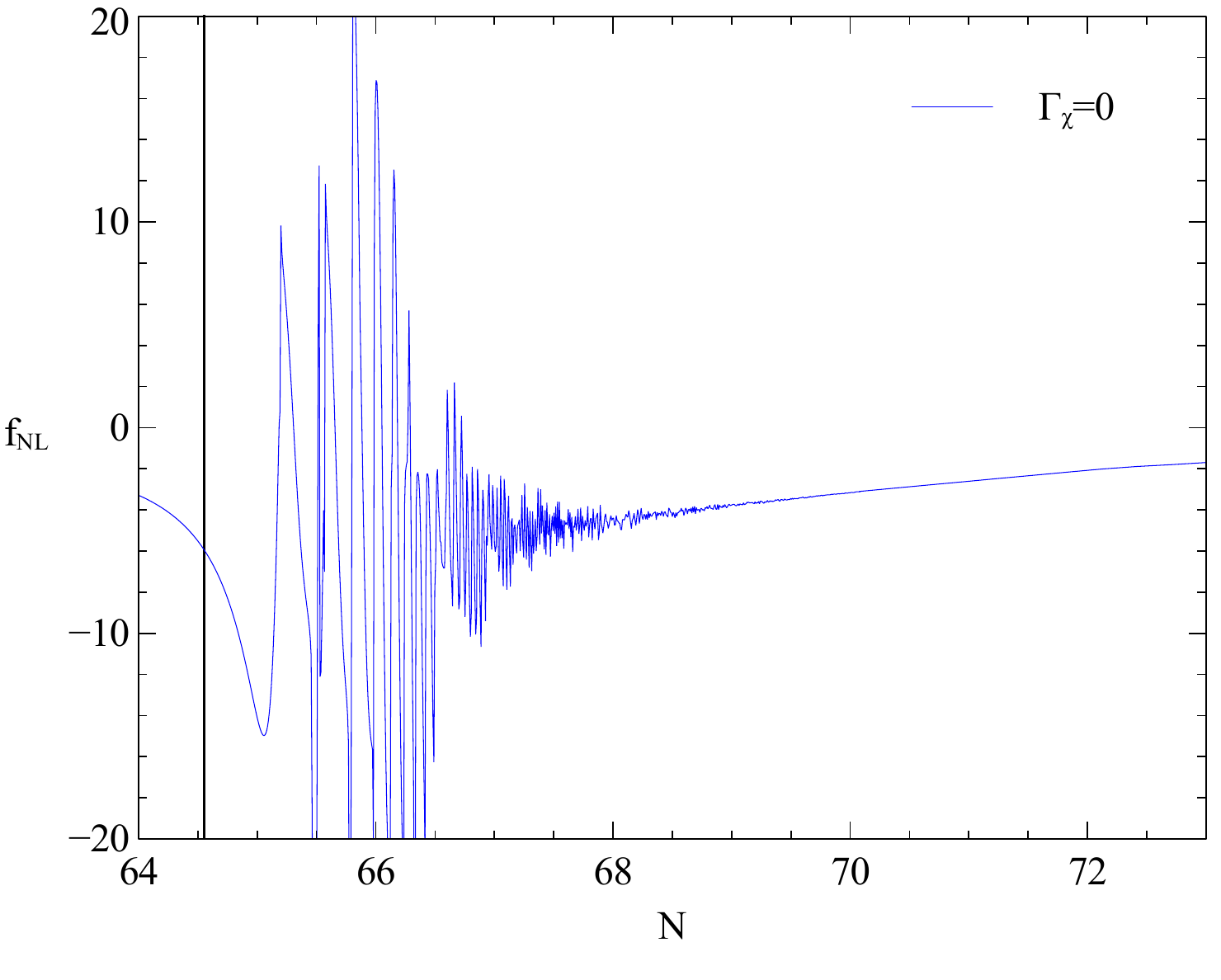}
	\end{tabular}	
	\caption{Potential: $W(\vp,\chi)=W_0\chi^2e^{-\lambda\vp^2}$. \textit{Left panel}: The evolution of the background fields for $\lambda=0.05$, $\vps=10^{-3}\Mp$ and $\chis=16.0\Mp$.  \textit{Right panel}: $\fnl$ as a function of $N$ (with $\Gchi=0$) after inflation has ended. The parameters used are: $\lambda=0.06$, $\vps=10^{-3}\Mp$ and $\chis=16.0\Mp$. In both panels, the solid vertical (black) line denotes the end of inflation, $N_{\rm e}$.}
	\label{fig:fNL-N-quadExpNoGamma}
\end{figure*}
It is useful to first consider the evolution of the fiducial background fields and $\fnl$ in the limit of no reheating, i.e., $\Gchi=0$. We set $\lambda=0.06$, $\vps=10^{-3}\Mp$ and $\chis=16.0\Mp$. With this choice of parameters, a large $\fnl$ is still present as slow--roll breaks down (at $N\sim64.5$). However, since no limiting trajectory is available, $\fnl$ continues to evolve: as $\chi$ reaches the minimum of the potential, it undergoes damped oscillations about $\chi=0$, which induces large oscillations in $\fnl$. The $\vp$ field remains in slow--roll after inflation has ended and throughout the period of oscillations of $\chi$ and continues to evolve towards ever increasing values. We note that if the potential is sufficiently steep in the $\vp$ direction, i.e., large $\lambda$, then the slow--roll conditions for $\vp$ may be violated. Such large values of $\lambda$ tilt the initial trajectory away from the axis of inflaton and hence a large $\fnl$ cannot be generated. As such, we do not consider such regions of parameter space. In Fig.~\ref{fig:fNL-N-quadExpNoGamma} we show the late time evolution of the scalar fields and $\fnl$. The `spikes' in the oscillations of $\fnl$ correspond to the $\chi$ field changing direction at the maximum of its oscillation. Fluctuations between neighbouring trajectories in the bundle source $\fnl$. These trajectories continue to diverge in the $\vp$ direction due to the geometry of the potential in this region and so $\fnl$ continues to evolve, decaying towards zero. If the evolution of $\fnl$ were followed indefinitely with $\Gchi=0$, we should expect it to settle to $\fnlfinal\approx0$. In order to explain this, we need to examine each derivative term contributing to the expression for $\fnl$, Eq.~(\ref{eq:fnl}). We begin by considering the slow--roll solution for $\vp$ and the $\etapp$ slow--roll parameter,
\be
\label{eq:quadExpSRvp}
\vp=\vps e^{2\lambda N}\,, \quad\quad \etapp=2\lambda\left[2\lambda\vps^2 e^{4\lambda N}-1\right]\,,
\ee
which shows that trajectories will continue to evolve indefinitely in the $\vp$ direction if $\Gchi=0$. This slow--roll solution for $\vp$ is shown against the exact numerical solution in the left panel of Fig.~\ref{fig:fNL-N-quadExpNoGamma}.

Now, inspecting the individual derivative terms ($N_I=\partial N/(\partial \vp^I_*)$ etc) contributing to $\fnl$, we find that $\Nchi$ remains practically constant, $\Nchi\approx(2\ecstar)^{-1/2}$, throughout the entire inflationary and post--inflationary phase, acquiring this value when the fields leave the horizon. To explain this, we assume that $H$ is monotonic in time, enabling us to re--write Eq.~(\ref{eq:numEfoldsDef}) as
\be 
\label{eq:numEfoldsDef2}
N=\int^{c}_{*} \frac{{\rm d}H^2}{2\dot{H}}\,.
\ee
Taking the derivative with respect to $\chis$ we find
\be 
\label{eq:dNdchis1}
\Nchi=\left(\frac{1}{2\dot{H}}\frac{\p H^2}{\p\chis}\right)_* +  \int^{c}_{*}  \frac{\p}{\p\chis}\left( \frac{1}{2\dot{H}}\right)_H {\rm d}H^2  \,,
\ee
where the derivative inside the integral is computed by holding $H$ constant. The derivative at the boundary $c$ vanishes, since by definition the surface $c$ corresponds to one of constant $H$. Using the fact that the fields are in slow--roll at horizon exit, the first term on the RHS of Eq.~(\ref{eq:dNdchis1}) reduces to $(2\ecstar)^{-1/2}$. Then, to explain why $\Nchi$ remains constant at this value requires arguing that the integral term in Eq.~(\ref{eq:dNdchis1}) is negligible, i.e., after perturbing $\chis$, surfaces of constant $\dot{H}$ must coincide with surfaces of constant $H$. This is indeed the case if a hierarchy of kinetic energies exists between the fields at horizon crossing, i.e., $|\dot{\chi}_*|\gg|\dot{\vp}_*|$. Since the kinetic terms are canonical, the fields follow the gradient of the potential, and as they are in slow--roll at horizon exit, this hierarchy implies $|W_\chi|_*\gg|W_\vp|_*$. If this is the case, the dependence of $\dot{H}$ on $\chis$ is rapidly washed out, and the two--dimensional bundle in the $\chi$ direction (holding $\vps$ fixed) degenerates to a caustic. We have found that the condition $|W_\chi|_*\gg|W_\vp|_*$ is sufficient to guarantee that the integrand of Eq.~(\ref{eq:dNdchis1}) is always small from horizon crossing until oscillations of $\chi$ begin. During the oscillatory phase, the integrand oscillates about zero with an amplitude that decays with the Hubble expansion, and when integrated over many oscillations, the net result is a negligible correction to $\Nchi$. By the same argument, $\Nchichi$ remains roughly constant at $\Nchichi\approx 1-(\etacc/2\ec)_*$, which, for this particular potential is independent of $\lambda$ and the field values at horizon crossing, $\Nchichi\approx\frac12$. 

Furthermore, we find that the following approximate scaling relations hold to a remarkable accuracy throughout the entire inflationary and post--inflationary evolution:
\bea
\Nvpvp&\approx&\frac{\Nvp}{\vps}\,, \label{eq:derivs_scale} \\
\Nchivp&\approx&4\lambda \Nvp\Nchi  \approx\frac{4\lambda}{\sqrt{2\ecstar}}\Nvp \,.\label{eq:derivs_scale2}
\eea
The scaling relation between $\Nvpvp$ and $\Nvp$ was first derived in~\cite{elliston:2011} by considering a first order Taylor expansion about a `ridge', situated at $\vp=0$, of a generic potential. 
\begin{figure*}[t]
	\begin{tabular}{cc}
		\includegraphics[width=0.48\linewidth]{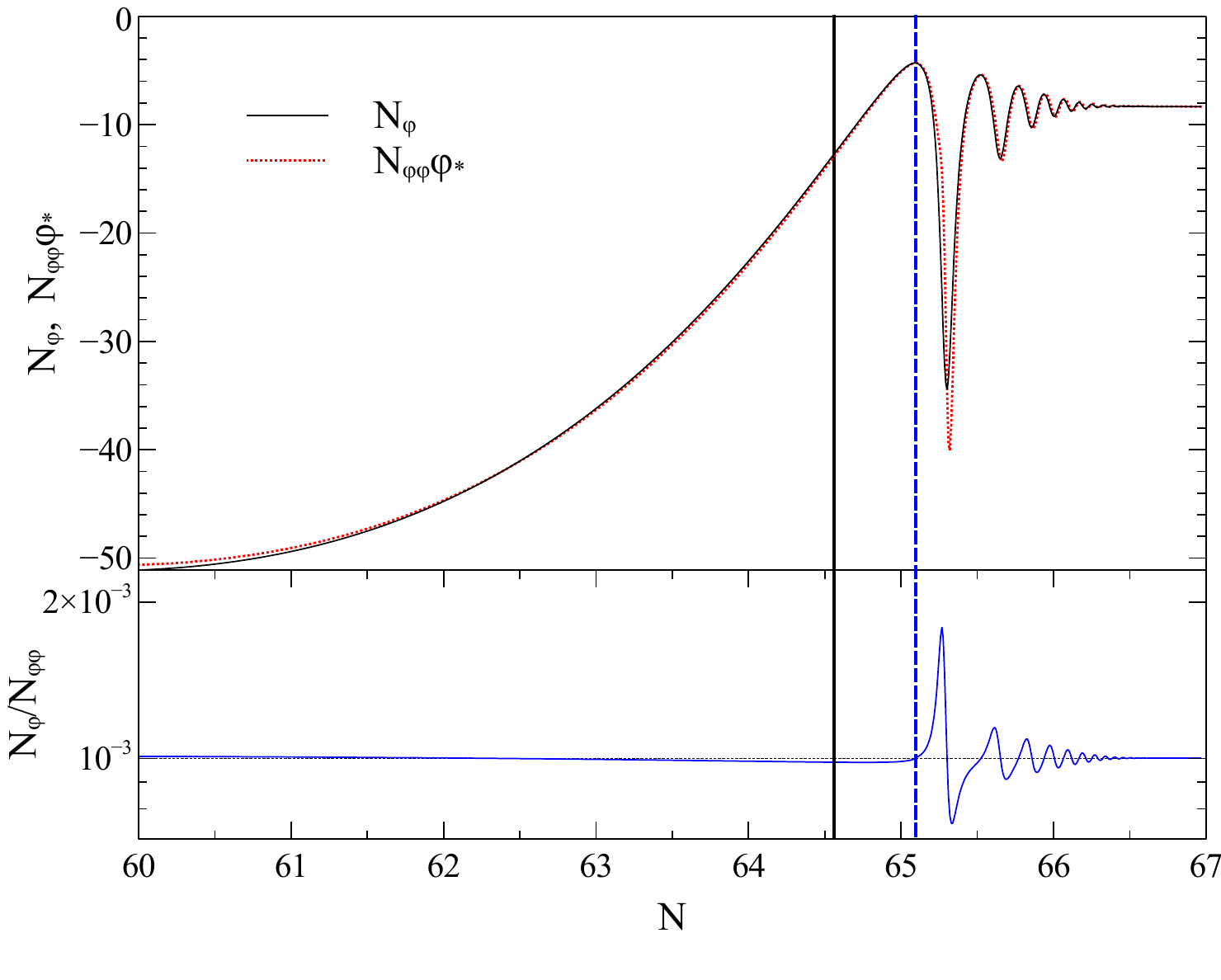}
		\includegraphics[width=0.48\linewidth]{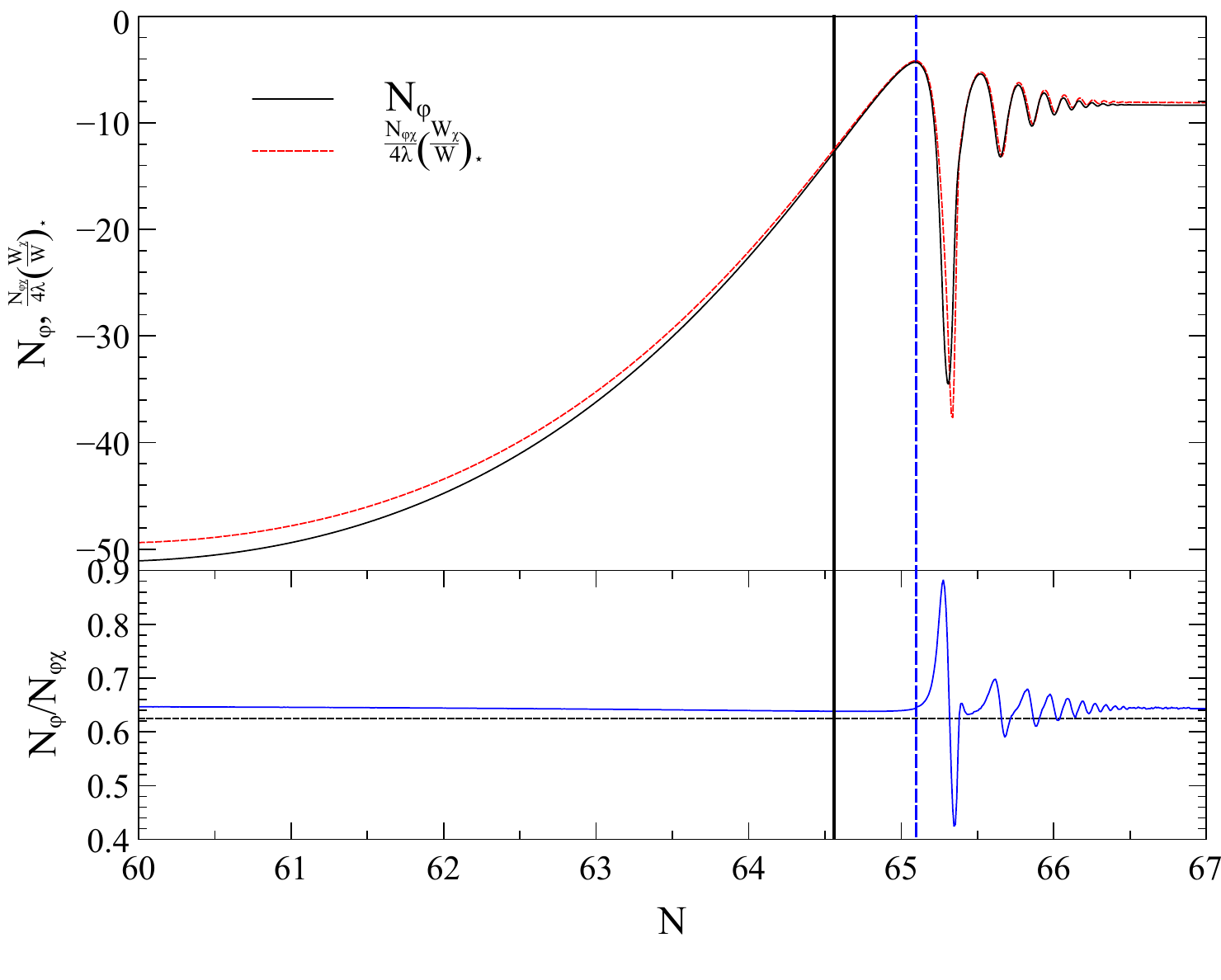} 
	\end{tabular}	
\caption{Potential: $W(\vp,\chi)=W_0\chi^2e^{-\lambda\vp^2}$. Numerical verification of the scaling relations Eqs.~(\ref{eq:derivs_scale}) and~(\ref{eq:derivs_scale2}). \textit{Left panel}: Evolution of the derivatives $\Nvpvp$ and $\Nvp$. The horizontal dashed line in the lower panel represents the value of $\vps$, the constant of proportionality between $\Nvpvp$ and $\Nvp$.  \textit{Right panel}: Evolution of the derivatives $\Nchivp$ and $\Nvp$. The horizontal dashed line in the lower panel represents the value $\frac{1}{4\lambda}(W_\chi/W)_*=\frac{1}{4\lambda}(2\ecstar)^{1/2}$, the constant of proportionality between $\Nchivp$ and $\Nvp$. We show evolution of the derivatives for the last few $e$--folds of inflation, up until $\zeta$ has become conserved at the completion of reheating. We see small departures from scaling at the start of reheating as $\chi$ oscillates about its minimum, but as $\chi$ settles down, the scaling behaviour is quickly recovered.  In both panels, the parameters used are: $\lambda=0.05$, $\vps=10^{-3}\Mp$, $\chis=16.0\Mp$ and $\Gchi=\sqrt{10^{-1}W_0}\Mp$. The solid vertical (black) line denotes the end of inflation, $N_{\rm e}$, and the dashed vertical (blue) line denotes the start of reheating, $N_{\rm r}$. The Hubble rate at the start of reheating is $H_r\approx\sqrt{7\times10^{-2}W_0}\Mp$.}
	\label{fig:derivs-N-quadExp-scaling}
\end{figure*}
Assuming the slow--roll conditions, the same analysis applies to the model we study here as long as the potential remains well approximated by $W\approx W_0\chi^2(1-\lambda\vp^2)$, i.e., higher order terms in $\lambda\vp^2$ remain small. This requires $\vp\ll\mathcal{O}(\lambda^{-1/2})$. In this regime, $\vp$ grows exponentially with $H$ as the bundle of trajectories rolls off the ridge: $\vp=\vps e^{\alpha(H_*^2-H^2)}$, $\alpha=3\lambda/2W_0$. A short calculation reveals 
\be
\label{eq:quadExp_Nvp_slowroll}
\Nvp \approx -3\beta H^{2} \vp_* \left(\frac{\vp}{\vp_*}\right)^2\,,
\ee
where $\beta$ is some model--dependent constant. We refer the reader to~\cite{elliston:2011} where the complete derivation is presented. Taking $\frac{\partial}{\partial\vps}$ (on surfaces of constant $H$) on both sides of Eq.~(\ref{eq:quadExp_Nvp_slowroll}) gives Eq.~(\ref{eq:derivs_scale}). Similarly, taking the derivative with respect to $\chis$ and using the $\vp$ slow--roll solution Eq.~(\ref{eq:quadExpSRvp}) gives Eq.~(\ref{eq:derivs_scale2}).

We show evolution of the $\Nvp$, $\Nvpvp$ and $\Nchivp$ derivatives before and \textit{after} inflation in Fig.~\ref{fig:derivs-N-quadExp-scaling}, which clearly illustrates the scaling behaviour captured in Eqs.~(\ref{eq:derivs_scale}) and~(\ref{eq:derivs_scale2}). Remarkably, not only does this scaling behaviour hold after inflation has ended, but it also holds during reheating. Indeed, we find that it remains an excellent approximation across the entire range of $\Gchi$ that are within our numerical capabilities, including $\Gchi=0$. 

The derivation of these scaling relations as sketched above relies on a number of approximations, including slow--roll. The sub--dominant field $\vp$ always remains slowly rolling, however $\chi$ necessarily does not. This does not seem to violate Eqs.~(\ref{eq:derivs_scale}) and~(\ref{eq:derivs_scale2}), suggesting that validity of these relations are more reliant on $\vp$ being a linear function of $\vps$, and that $\vp$ grows exponentially as the bundle slides off the ridge. As mentioned above, these conditions will break down when $\vp\sim\mathcal{O}(\lambda^{-1/2})$. Then, using $\vp\sim\lambda^{-1/2}$ in Eq.~(\ref{eq:quadExpSRvp}) we may very roughly estimate how many $e$--folds we expect the scaling relations to remain valid: $N\sim\frac{1}{2\lambda}{\rm ln}\, (\lambda^{-1/2}/\vps)$. For example, for $\lambda=0.05$ and $\vps=10^{-3}$ we have $N\sim85$. 

We now return to the expression for $\fnl$, Eq.~(\ref{eq:fnl}). The $\Nchichi$ and $\Nchivp$ derivatives are in fact negligible compared to $\Nvpvp$ and so can be safely neglected. Making use of the approximations discussed above, we may write $\fnl$ solely in terms of $\Nvp$:
\be
\label{eq:fnlApprox1}
\fnl\approx\frac{5}{6|\vps|}\frac{\Nvp^3}{[\Nvp^2+g_*^2]^2}\,,
\ee
where  $g_*\equiv N_{\chi}\approx (2\ecstar)^{-1/2}$.  The asymptotic behaviour of $\fnl$ is clear: trajectories in the bundle continue to diverge away from one another in the $\vp$ direction according to Eq.~(\ref{eq:quadExpSRvp}), which continuously sources $\Nvp$, making it grow increasingly more negative. Hence, in the limit that $\Nvp\rightarrow-\infty$ we expect $\fnlfinal\rightarrow0$, which is what is observed in the right panel of Fig.~\ref{fig:fNL-N-quadExpNoGamma}, justifying our previous claims. The sign of $\Nvp$ can be argued from the geometry of the potential: diverging trajectories source negative $\Nvp$ (as the sign of Eq.~(\ref{eq:quadExp_Nvp_slowroll}) indicates), whilst converging trajectories source positive $\Nvp$~\cite{elliston:2011}. \\

With the limiting case $\Gchi=0$ understood, we now move on to explore the dependence of $\fnlfinal$ on $\Gchi$, keeping the same parameter choice $\lambda=0.06$, $\vps=10^{-3}\Mp$ and $\chis=16.0\Mp$. We express the decay rates in terms of the overall potential normalisation, $W_0$. Whilst its value sets the scale of inflation and determines the amplitude of the primordial power spectrum and hence is constrained, it does not affect the statistics of $\zeta$ and so we leave $W_0$ as a free parameter. Where applicable, we also give the value of the Hubble rate at the start of reheating, $H_r$, in units of $W_0$ so a direct comparison between the expansion and decay rate can be made.
\begin{figure*}[t]
	\begin{tabular}{cc}
		\includegraphics[width=0.48\linewidth]{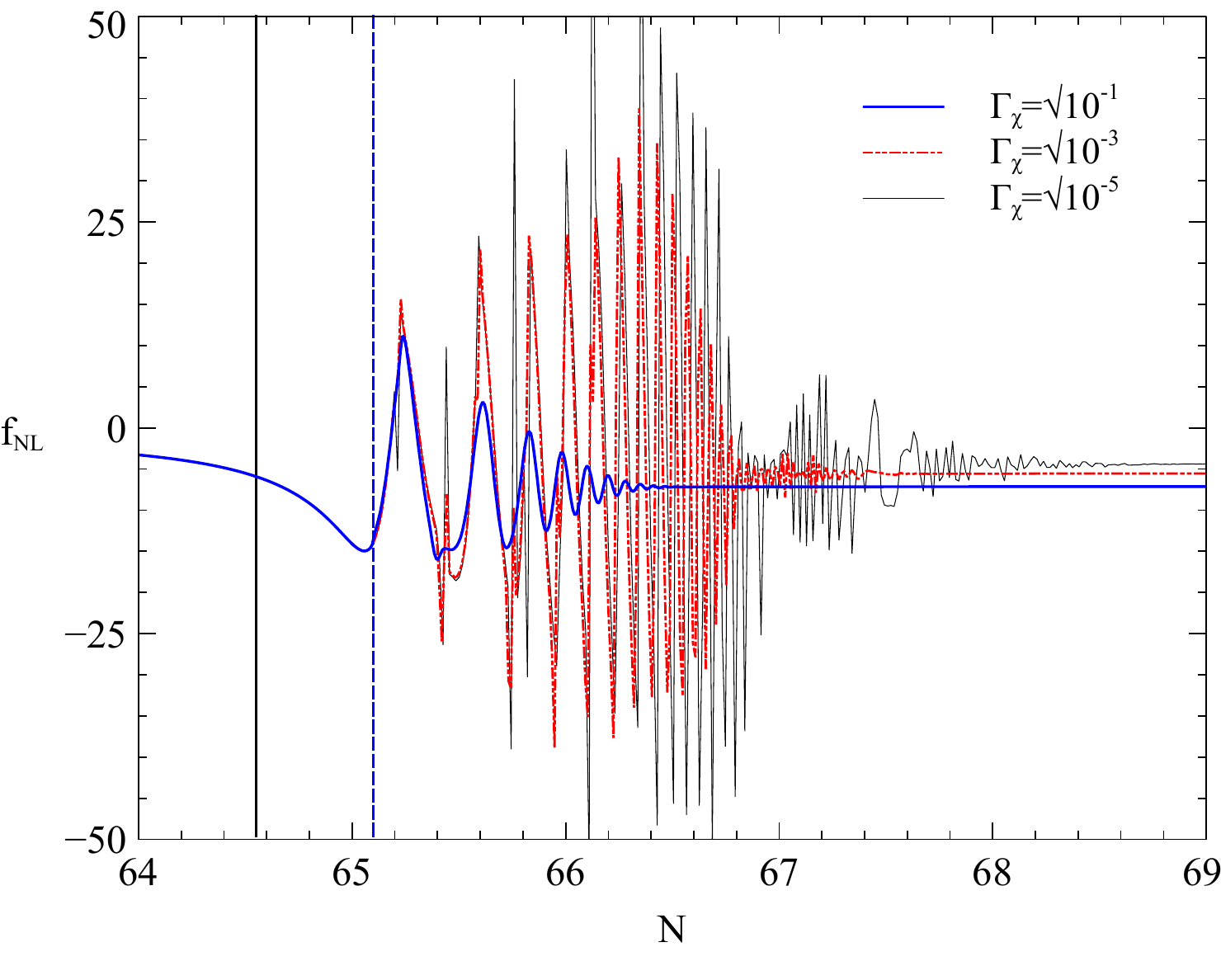} 	&
		\includegraphics[width=0.48\linewidth]{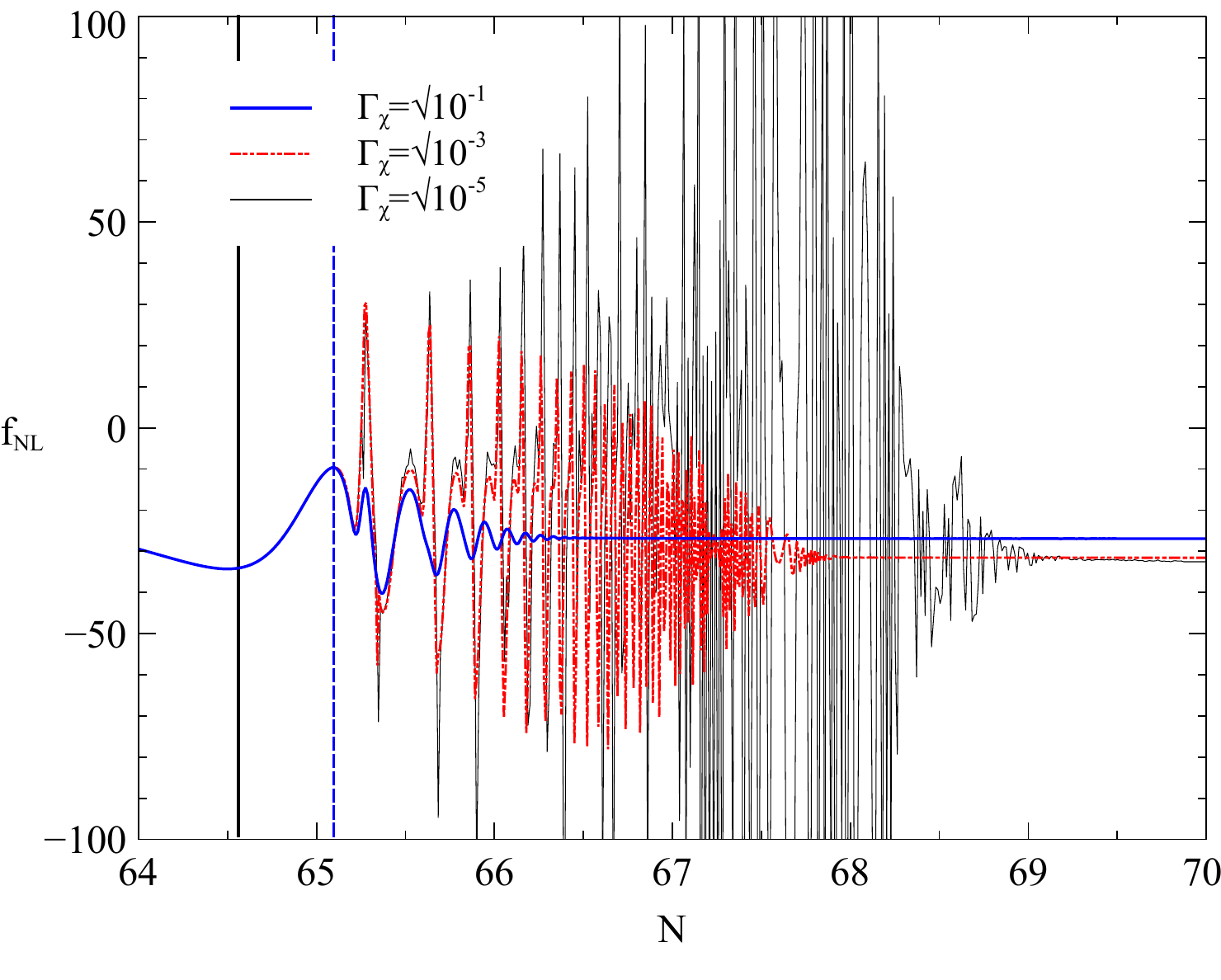} 
	\end{tabular}	
	\caption{Potential: $W(\vp,\chi)=W_0\chi^2e^{-\lambda\vp^2}$. We show the evolution of $\fnl$ during reheating for various decay rates $\Gchi$, which are in units of $\sqrt{W_0}\Mp$. In both panels, the solid vertical (black) line denotes the end of inflation, $N_{\rm e}$, and the dashed vertical (blue) line denotes the start of reheating, $N_{\rm r}$. \textit{Left Panel}: The parameters used are: $\lambda=0.06$, $\vps=10^{-3}\Mp$ and $\chis=16.0\Mp$.  The Hubble rate at the start of reheating is $H_r\approx\sqrt{7\times10^{-2}W_0}\Mp$. \textit{Right Panel}: The parameters used are: $\lambda=0.05$, $\vps=10^{-3}\Mp$ and $\chis=16.0\Mp$. The Hubble rate at the start of reheating is $H_r\approx\sqrt{6\times10^{-2}W_0}\Mp$. }
	\label{fig:fNL-N-quadExp}
\end{figure*}

The switching on of the decay terms at the reheating surface sources the radiation density. As the $\chi$ field oscillates about its minimum, its kinetic energy is transferred to the radiation fluid, resulting in bursts of particle production. As radiation fills the universe, Hubble damping slows the motion of $\vp$ to a crawl and as we approach $\Omega_\gamma\sim1$, it asymptotes to a constant: $\vp(t\rightarrow\infty)\approx const$. Herein is the fundamental difference in the motion of $\vp$ when $\Gchi\neq0$ compared to $\Gchi=0$: as radiation comes to dominate, trajectories in the bundle cease to evolve. The bundle does not degenerate to a caustic as would be the case if the trajectories were naturally focussed by a region of the potential, but nonetheless this freezing of the $\vp$ field guarantees that $\zeta$ becomes conserved. This does not happen in the $\Gchi=0$ limit where the trajectories continue to diverge in the $\vp$ direction, always sourcing $\zeta$.

In the left panel of Fig.~\ref{fig:fNL-N-quadExp} we show the final stages in the evolution of $\fnl$ as a function of $N$ for various decay rates $\Gchi$. Most importantly, we see that reheating does \textit{not} damp out $\fnl$ to zero. We interpret the fine details of the plot as follows: At the end of inflation ($N_{\rm e}=64.56$) a large, negative $\fnl$ is still present, and just before reheating begins\footnote{Here, we use the terminology `start of reheating' to refer to the time when the fiducial background trajectory, emanating from $\{\chis,\vps\}$, crosses $\chi_0$ for the first time.} ($N_{\rm r}=65.10$) $\fnl$ is growing increasingly more negative.  We see that as the decay rate $\Gchi$ is increased from zero, $|\fnlfinal|$ freezes out to larger values. In another example where $\fnl$ is decaying toward zero as reheating begins, the effect of increasing the decay rate from zero is to freeze out $|\fnlfinal|$ to smaller values. This is shown in the right panel of Fig.~\ref{fig:fNL-N-quadExp}.

This opposite dependence of $|\fnlfinal|$ on $\Gchi$ for $\lambda=0.05$ and $\lambda=0.06$ is a consequence of the non--trivial dependence of $\fnl$ on $\Nvp$. Let us begin by considering the splitting
\be 
\label{eq:Nsplit}
N=\int^{c}_{*} \frac{{\rm d}H^2}{2\dot{H}} = \int^{r}_* \frac{{\rm d}H^2}{2\dot{H}} + \int^{c}_{r} \frac{{\rm d}H^2}{2\dot{H}} = N_0+N_1\,.
\ee
Here $N_0$ is the number of $e$--foldings from horizon crossing ($t_*$) up to the start of reheating ($t_r$) and $N_1$ is the number of $e$--foldings from the start of reheating up to radiation domination ($t_c$). Firstly, it is important to appreciate that $N_0$ contains contributions not only from the slow--roll inflationary phase, but also from the non--negligible post--inflation/pre--reheating evolution, that must be accounted for. Whilst the standard methods (see eg. Refs.~\cite{Vernizzi2006NonGaussianities,Choi2007Spectral,Byrnes2008Conditions,Meyers2011NonGaussianities}) may be used to compute the derivatives ($N_{,I}$ etc) of the slow--roll contribution to $N_0$, derivatives of the remaining non--slow--roll contribution to $N_0$ cannot be calculated explicitly. Secondly, $N_0$ does not contain any dependence on the reheating process. Since we are interested here in studying the effects of reheating on $\fnlfinal$, we compute $N_0$ and its derivatives numerically and focus on trying to understand the correction $N_1$, which contains all the dependence on $\Gchi$. 

\begin{figure*}[t]
	\begin{tabular}{cc}
		\includegraphics[width=0.48\linewidth]{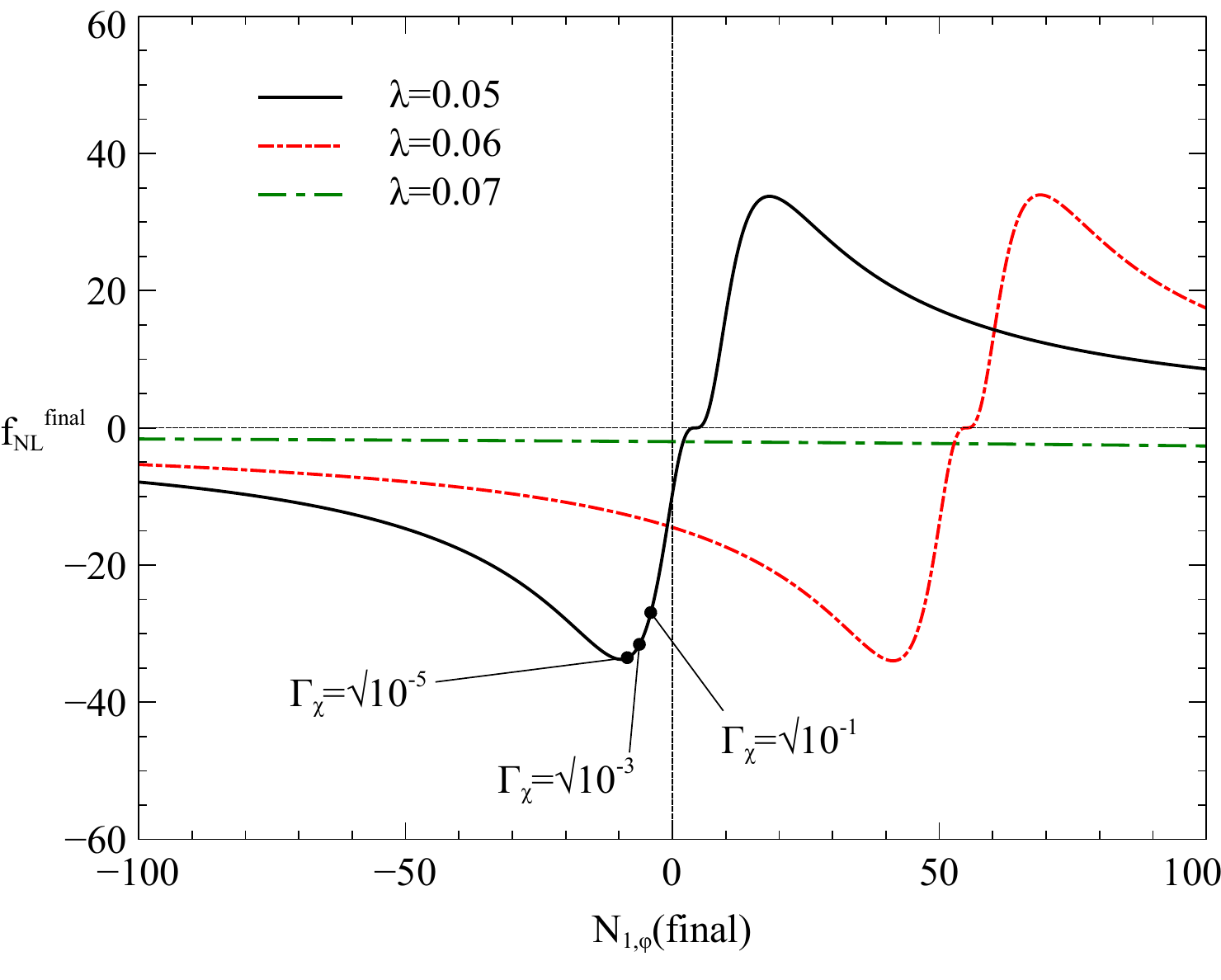}
		\includegraphics[width=0.48\linewidth]{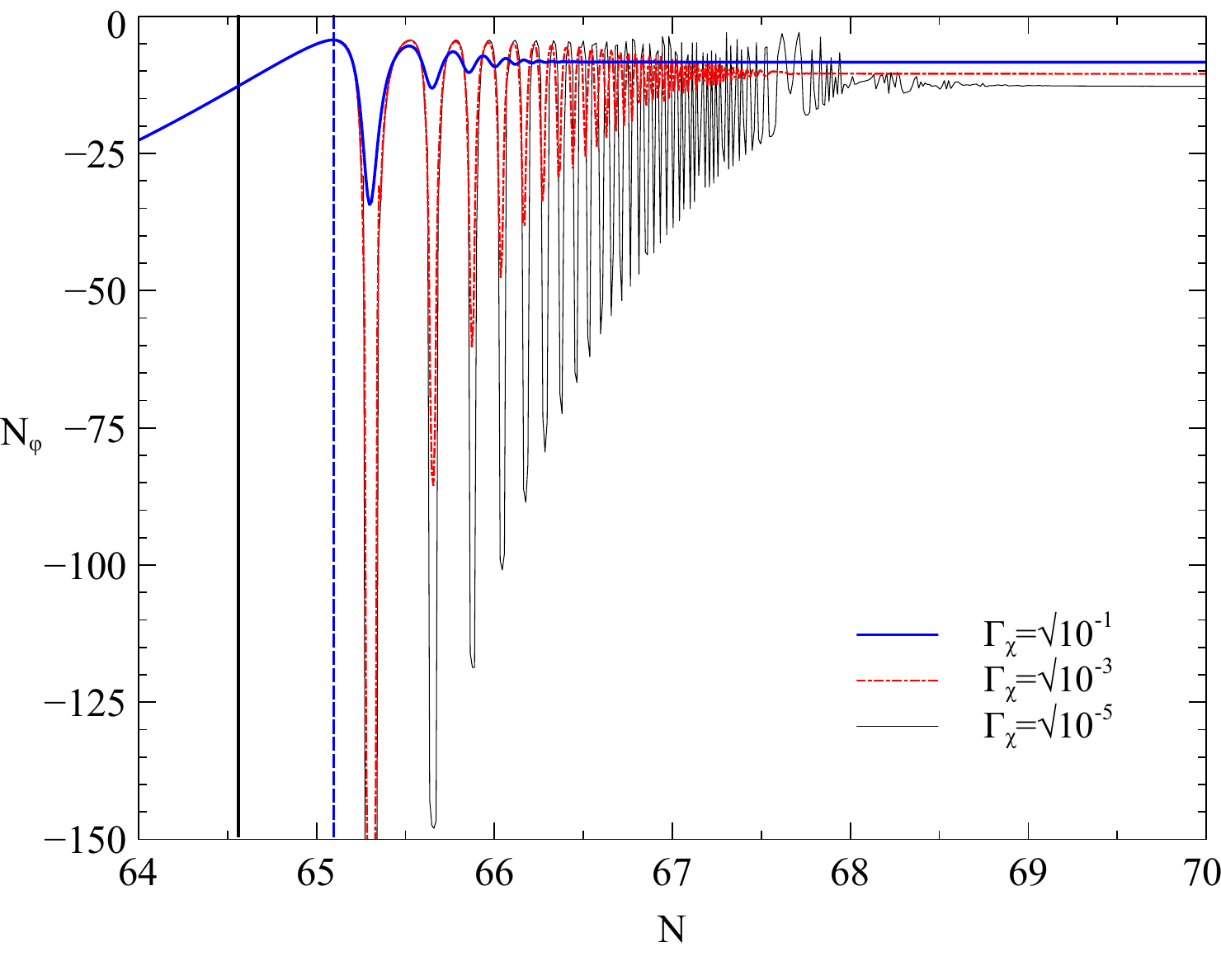}
	\end{tabular}	
	\caption{Potential: $W(\vp,\chi)=W_0\chi^2e^{-\lambda\vp^2}$. \textit{Left Panel}: The algebraic function $\fnlfinal$ as a function of the final value of the correction $N_{1,\vp}$, Eq.~(\ref{eq:fnlApprox2}). We label the positions along the $\lambda=0.05$ curve which correspond to the decay rates given in the right panel. \textit{Right Panel}: The evolution of the derivative $\Nvp=N_{0,\vp}+N_{1,\vp}$ for the same decay rates as Fig.~\ref{fig:fNL-N-quadExp}, for $\lambda=0.05$. All decay rates are in units of $\sqrt{W_0}\Mp$. The solid vertical (black) line denotes the end of inflation, $N_{\rm e}$, and the dashed vertical (blue) line denotes the start of reheating, $N_{\rm r}$. The Hubble rate at the start of reheating is $H_r\approx\sqrt{7\times10^{-2}W_0}\Mp$.}
	\label{fig:quadexpfNLanal}
\end{figure*}

For the derivative of the correction $N_1$ with respect to $\vps$ we need only consider the term
\be
\label{eq:phiDeriv}
N_{1,\vp}=\int^{c}_{r} \frac{\p}{\p\vps} \left(\frac{1}{2\dot{H}}\right)_H {\rm d}H^2   \,,
\ee
since the derivative at the boundary at $r$ cancels with the $N_0$ contribution and derivative at the boundary $c$ vanishes since $c$ is defined as a surface of constant $H$. Since $\dot{H}$ is a function of $\dot{\chi}(t)$, $\dot{\vp}(t)$ and $\rho_\gamma(t)$, all of which depend on $\vps$, this integral cannot be performed analytically beyond slow--roll. However, we can make progress by using our results, $\Nchi\approx(2\ecstar)^{-1/2}$, $\{\Nchivp\,,\,\Nchichi\}<<\Nvpvp$ and $\Nvpvp\sim\Nvp/\vps$ which also hold during reheating. Then, using the fact that during reheating $N_{1,\chi}\approx0$, and taking the time $t_c$ to be deep in the radiation dominated era such that $N_{1,\vp}=const$, Eq.~(\ref{eq:fnlApprox1}) becomes
\be
\label{eq:fnlApprox2}
\fnlfinal \approx \frac{5}{6|\vps|}\frac{ (N_{0,\vp}+N_{1,\vp})^3  }{[(N_{0,\vp}+N_{1,\vp})^2+g_*^2]^2}\,.
\ee
We plot this \textit{algebraic} function, $\fnlfinal$ against $N_{1,\vp}$, in the left panel of Fig.~\ref{fig:quadexpfNLanal} for three different choices of the potential parameter $\lambda=\{0.05\,,0.06\,,0.07\}$, with the same field values at horizon crossing $\vps=10^{-3}\Mp$ and $\chis=16.0\Mp$. Changing $\lambda$ obviously changes $g_*$ and modifies the evolution of the bundle, changing $N_{0,\vp}$. In the right panel of Fig.~\ref{fig:quadexpfNLanal} we show the evolution of $\Nvp$ for various decay rates with $\lambda=0.05$. The final values of $N_{1,\vp}({\rm final})=\Nvp({\rm final})-N_{0,\vp}$ are marked on the corresponding curve in the left panel of Fig.~\ref{fig:quadexpfNLanal}.
Only the $N_{1,\vp}\leq0$ region of Eq.~(\ref{eq:fnlApprox2}) is physical: we have already argued that diverging trajectories can only generate negative $N_{1,\vp}$, and we have confirmed this numerically. As can be seen from the left panel of Fig.~\ref{fig:quadexpfNLanal}, Eq.~(\ref{eq:fnlApprox2}) has three stationary points at finite $N_{1,\vp}$: 
\be
-N_{0,\vp}\,, \quad -N_{0,\vp}\pm\sqrt{3}g_*\,.
\ee
The $N_{1,\vp}=-N_{0,\vp}$ root is an inflection point where $\fnlfinal=0$. The $N_{1,\vp}=-N_{0,\vp}+\sqrt{3}g_*$ root is a local maximum where $\fnlfinal$ would be always positive and so is not physical. The minimum at $N_{1,\vp}=-N_{0,\vp}-\sqrt{3}g_*$ however is physical and bounds the maximum value of $|\fnlfinal|$ when Eq.~(\ref{eq:fnlApprox2}) has a minimum at negative $N_{1,\vp}$:
\be
\label{eq:fnlatmin}
|\fnlfinal|_{\rm max}\approx\frac{1}{|g_*\vps|}\sqrt{\frac{75}{1024}}\,,  \quad {\rm for}  \quad N_{0,\vp}+\sqrt{3}g_*>0\,.
\ee
A minimum at negative $N_{1,\vp}$ is clearly seen in the left panel of Fig.~\ref{fig:quadexpfNLanal} for $\lambda=0.05$. If on the other hand, the minimum exists at positive $N_{1,\vp}$, (i.e., $N_{0,\vp}+\sqrt{3}g_*<0$) then the maximum value of $|\fnlfinal|$ is instead bounded by its value at the start of reheating:
\be
\label{eq:fnlatReheat}
|\fnlfinal|_{\rm max}\approx|\fnl(t_r)|\approx\frac{5}{6|\vps|}\frac{ N_{0,\vp}^2 }{[N_{0,\vp}^2+g_*^2]^2}\,, \quad {\rm for}  \quad N_{0,\vp}+\sqrt{3}g_*<0\,.
\ee
This is the case for the $\lambda=0.06$ and $\lambda=0.07$ models shown in the left panel of Fig.~\ref{fig:quadexpfNLanal}. These bounds are independent of the decay rate $\Gchi$. Furthermore, the bound Eq.~(\ref{eq:fnlatmin}) is written solely in terms quantities evaluated at horizon crossing, and hence may be computed without explicitly knowing the full non--linear evolution of the bundle during the reheating process. Whether this maximum value, $|\fnlfinal|_{\rm max}$, is obtained at the end of reheating is of course dependent on $\Gchi$. Formally, the lower bound for $\fnlfinal$ (which is approached as $\Gchi\rightarrow0$) would be ${\fnlfinal}_{\rm min}=0$.

The existence of a minimum of Eq.~(\ref{eq:fnlApprox2}) at negative $N_{1,\vp}$ for $\lambda=0.05$ explains the seemingly opposite dependence of $\fnlfinal$ on $\Gchi$ compared to $\lambda=0.06$ where the minimum exists at positive $N_{1,\vp}$: as $\Gchi$ is increased from zero (where $N_{1,\vp}\rightarrow-\infty$), the time taken for reheating to complete is reduced and $\vp$ freezes out sooner, hence reducing the magnitude of $N_{1,\vp}$. For the $\lambda=0.05$ model, as $\Gchi$ is increased further, driving $N_{1,\vp}$ toward zero, the minimum of Eq.~(\ref{eq:fnlApprox2}) is encountered, past which point $|\fnlfinal|$ is reduced. For $\lambda=0.06$, increasing $\Gchi$ still drives $N_{1,\vp}$ toward zero, but this time $|\fnlfinal|$ is increased. 

For $\lambda=0.07$, the function Eq.~(\ref{eq:fnlApprox2}) is almost completely flat for $N_{1,\vp}<0$, which indicates that no matter how slowly or rapidly the universe is reheated, the value of $\fnl$ at the start of reheating will survive until completion. In the limit of instantaneous reheating, $\Gchi\rightarrow\infty$, $N_{1,\vp}\approx0$, and so $\fnlfinal\approx\fnl(t_r)$. This is only approximate since, as reheating does not begin on a hypersurface of constant density, there will be some small correction $N_{1,\vp}$. 

Another interesting observation is that $|\fnlfinal|$ (or more accurately the derivative $N_{1,\vp}$) is fairly insensitive to changing the decay rate by many orders of magnitude. For example, as can be seen from Table~\ref{tab:oneminstats}, $|\fnlfinal|$ changes by less than three units as the decay rate is increased from $\Gchi=\sqrt{10^{-5}W_0}\Mp$ to $\Gchi=\sqrt{10^{-1}W_0}\Mp$. We caution here that decay rate could, in principle, be many orders of magnitude weaker than the weakest decay rate studied here and still be consistent with the bound derived from BBN constraints, $\Gchi\gtrsim 4\times10^{-40}\Mp$. These tiny (but non--zero) values of $\Gchi$ are beyond our numerical capabilities: to compute the statistics of $\zeta$ at the completion of reheating requires integrating the field equations up until the universe is radiation dominated, which for such weak rates, can take $\mathcal{O}(30)$ $e$--folds. Substantial errors are accumulated if the field equations are integrated over such long periods of time, which in turn induces large errors in the computation of the $\Dn$ derivatives. For this reason, we only quote values of $\fnl$, $\nz$ and $r$ for decay rates for which we are confident that we have control over all sources of numerical error.

\subsection{Quartic minimum: $W(\vp,\chi)=W_0\chi^4e^{-\lambda\vp^2}$}\label{sec:oneminquartic}

We now repeat the same analysis, but with a quartic minimum in the $\chi$ direction. The background inflationary dynamics are similar to the $\chi^2 e^{-\lambda\vp^2}$ model as can be seen from the slow--roll solutions to the field equations:
\be
\label{eq:SRquarticExpsols}
\chi^2=\chi_*^2-8N\,, \quad \quad \vp=\vp_* e^{2\lambda N}\,.
\ee
The oscillatory dynamics about the minimum are somewhat different to that of the $\chi^2$ case however, due to the potential being much shallower around $\chi=0$, with steep sides away from the minimum. The oscillations of $\chi$ are not sinusoidal, but are instead given approximately by the elliptic function~\cite{Kofman1996Origin,Bassett2006Inflation} when $\Gchi=0$:
\be
\label{eq:quarticExpElipticalSols}
\chi(\tau)\approx \sqrt{\frac{3}{2\pi}}\Mp\frac{b}{\omega\tau}\, cn\left(\frac{\omega}{c}\tau,\,\frac{1}{\sqrt{2}}  \right)\,,
\ee
where $\tau$ is the conformal time ${\rm d}t=a(t){\rm d}\tau$. Here, $b\approx0.85$ is a numerical constant and $\omega$ is the effective frequency of the oscillations. The energy density of the coherently oscillating $\chi$ field decreases in the same way as a relativistic fluid, $\rho_\chi\sim a^{-4}$, behaving as radiation with a non--vanishing pressure $P_\chi\approx\frac13 \rho_\chi$. 

\begin{figure*}[t]
	\begin{tabular}{cc}
		\includegraphics[width=0.48\linewidth]{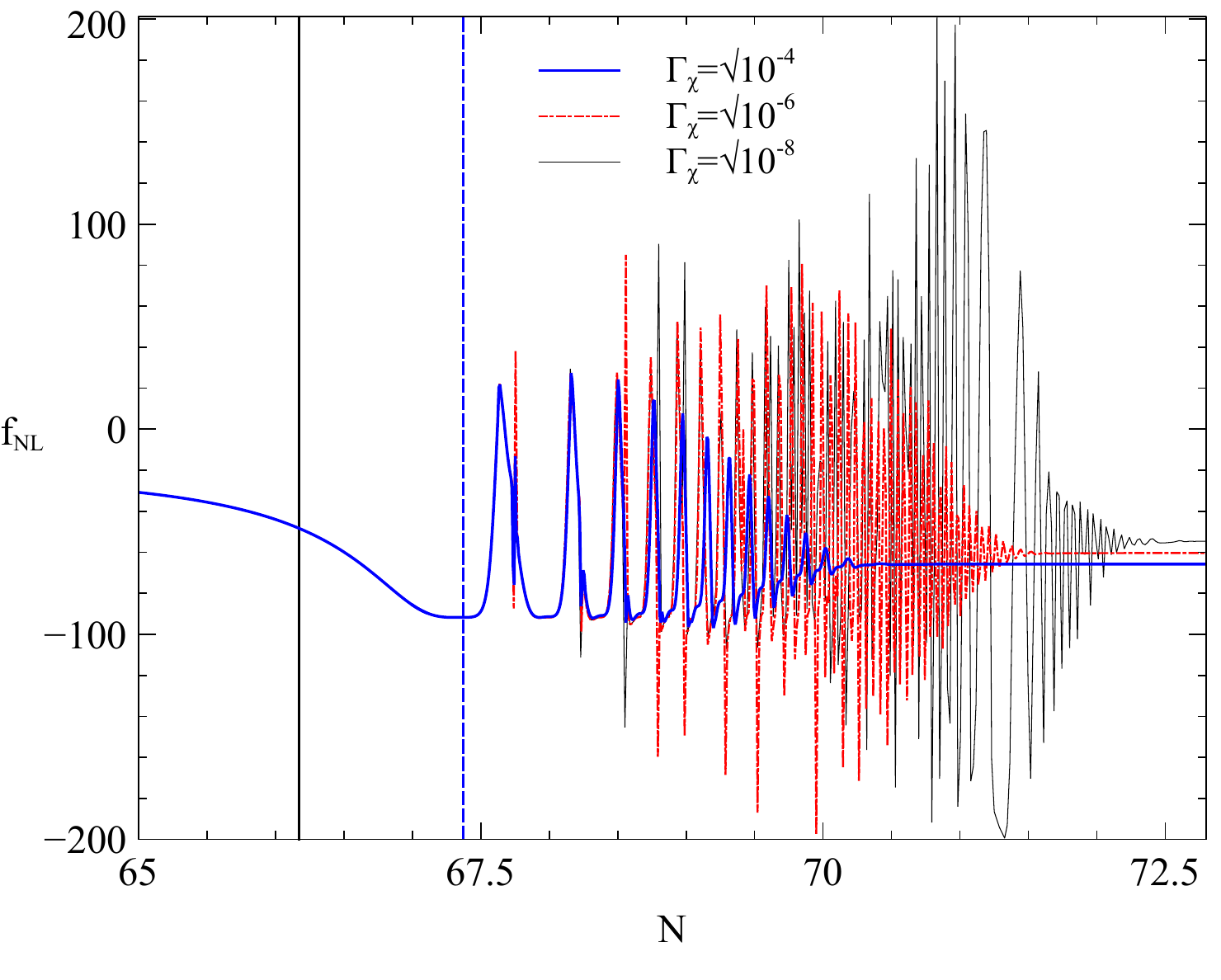}
		\includegraphics[width=0.48\linewidth]{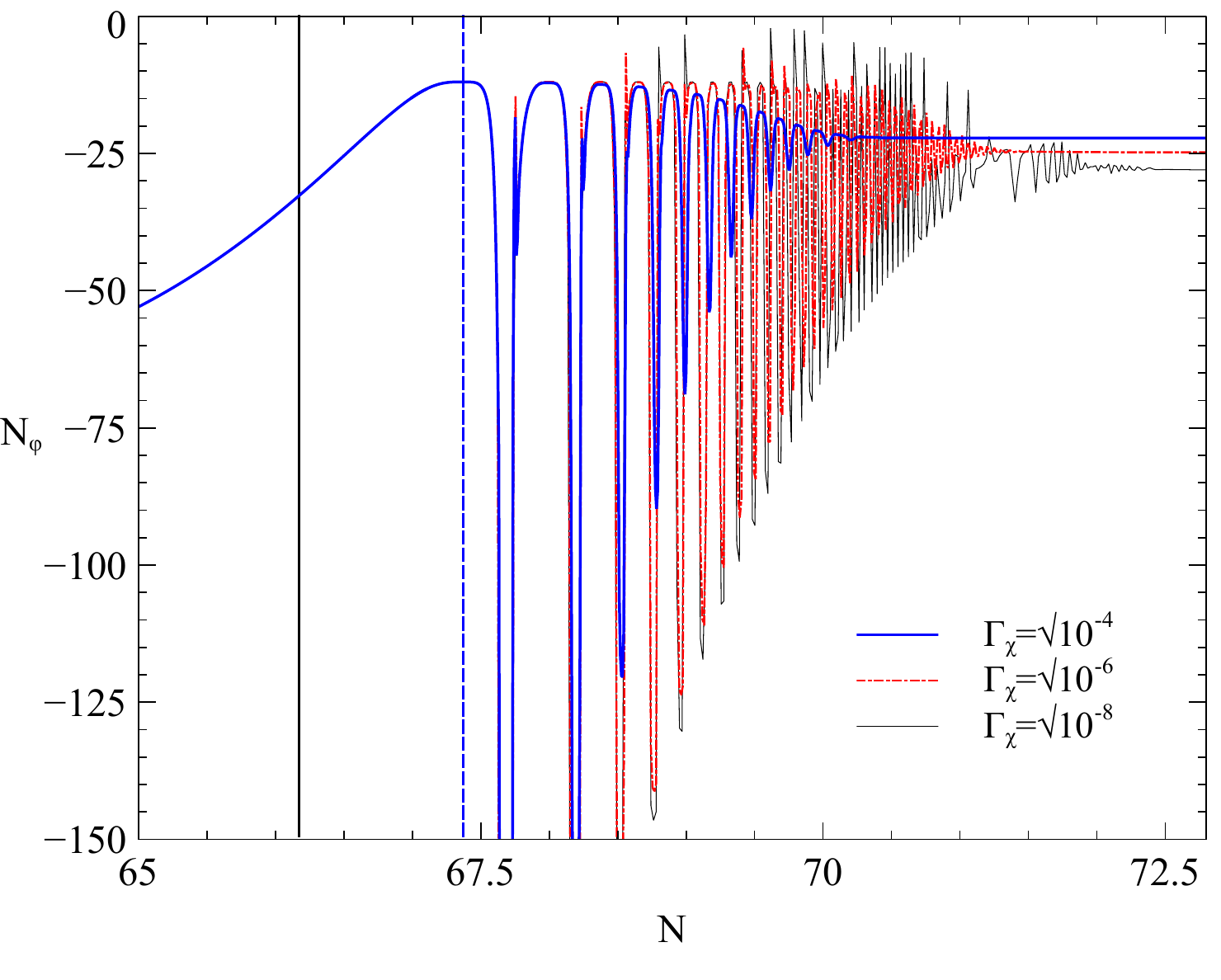}
	\end{tabular}	
	\caption{$W(\vp,\chi)=W_0\chi^4 e^{-\lambda\vp^2}$. The parameters used are: $\lambda=0.055$, $\vps=5\times10^{-4}\Mp$ and $\chis=23.0\Mp$. \textit{Left Panel}: The evolution of $\fnl$ during reheating for various decay rates $\Gchi$. \textit{Right Panel}: The evolution of the derivative $\Nvp$ during reheating for various decay rates $\Gchi$. All decay rates are in units of $\sqrt{W_0}\Mp$. In both panels, the solid vertical (black) line denotes the end of inflation, $N_{\rm e}$, and the dashed vertical (blue) line denotes the start of reheating, $N_{\rm r}$. The Hubble rate at the start of reheating is $H_r\approx\sqrt{10^{-1}W_0}\Mp$.}
	\label{fig:fNL-N-quarticExp}
\end{figure*}

Provided $\lambda$ is not too large, the $\vp$ field remains slowly rolling throughout the entire reheating phase. In the left and right panels of Fig.~\ref{fig:fNL-N-quarticExp} we show the final stages in the evolution of $\fnl$ and $\Nvp$ respectively as a function of $N$ for various decay rates $\Gchi$. We see that the qualitative dependence of $\fnlfinal$ on the decay rate is the same as for the $\chi^2 e^{-\lambda\vp^2}$ model, which may be explained by appealing to Eq.~(\ref{eq:fnlApprox2}). This implies that the shape of the minimum does not change the qualitative dependence of $\fnlfinal$ on the reheating process. Of course, as reheating proceeds, the shape of the $\chi$ minimum does not remain exactly quartic (or quadratic in the case of the previous model) due to the coupling with the $\vp$ field. We will comment more on this in Section~\ref{sec:discConclu}.
%

\subsection{Spectral index and tensor--to--scalar ratio}\label{sec:sindex_tensor}

For completeness, we also examine how sensitive the tensor--to--scalar ratio $r$, and spectral index $\nz$, are to the reheating phase for the models studied above. From Eqs.~(\ref{eq:scaltensR}) and (\ref{eq:spectrum}), we have that $r=8/\Mp^2(\sum_{I} N_{,I}^2)$. Recall that due to the hierarchy in magnitude between the scalar field kinetic energies $\dot{\vp}^2_*$ and $\dot{\chi}^2_*$ at horizon exit, we can approximate $g_*\equiv\Nchi\approx(2\ecstar)^{-1/2}$. For the region of parameter space of interest, as $\Gchi$ is decreased from infinity, the time taken for reheating to complete is increased and $\vp$ freezes out later, increasing the magnitude of $N_{1,\vp}$. Hence, the weaker the decay rate, the more suppressed the tensor--to--scalar ratio, and the following bound exists:
\be
\label{eq:scaltensR_bound}
r \leq \frac{8}{\Mp^2}\frac{1}{N_{0,\vp}^2 + g_*^2}\,.
\ee
This suppression of $r$ for weaker $\Gchi$ is illustrated in Table~\ref{tab:oneminstats}.

A similar bound also exists for the spectral index. Whilst it is a good approximation to neglect $\Nchivp$ in the expression for $\fnl$, one must be more careful in the expression for $\nz$: the hierarchy $|\dot{\vp}_*|\ll|\dot{\chi}_*|$ means that the sum in the numerator of Eq.~(\ref{eq:index}) generates terms of similar order.

\renewcommand*\arraystretch{1.2}
\begin{table}[h!]
\vspace{5pt}
    \begin{center}
        \subtable{
\begin{tabular}{c|c|c|c}
\multicolumn{4}{c}{$\chi^2$ minimum: $\fnl(t_e)=-5.93$, } \\
\multicolumn{4}{c}{$\nz(t_e)=0.763$, $r(t_e)=2.8\times10^{-4}$} \vspace{2mm}  \\
\hline
\hline
$\G_\chi$               &      $\fnlfinal$  &    $n_s^{\rm final}$   &  $r^{\rm final}$ \\
\hline
$\sqrt{10^{-5}}$   &	 	$-4.35$   &       $0.761$	    &        $2.4\times10^{-4}$  \\
$\sqrt{10^{-3}}$   &	 	$-5.54$   &       $0.762$	    &        $3.9\times10^{-4}$   \\
$\sqrt{10^{-1}}$   &	 	$-7.14$   &       $0.762$	    &        $6.3\times10^{-4}$  
                \end{tabular}\centering
            }
\hspace{10mm}
        \subtable{
\begin{tabular}{c|c|c|c}
\multicolumn{4}{c}{$\chi^4$ minimum: $\fnl(t_e)=-48.29$,} \\
\multicolumn{4}{c}{$n_s(t_e)=0.770$, $r(t_e)=7.2\times10^{-3}$} \vspace{2mm}  \\
\hline
\hline
$\Gchi$               &      $\fnlfinal$  &    $n_s^{\rm final}$   &  $r^{\rm final}$ \\
\hline
$\sqrt{10^{-8}}$   &	 	$-54.40$       &       $0.772$		    &        $9.7\times10^{-3}$  \\
$\sqrt{10^{-6}}$   &	 	$-60.32$       &       $0.778$		    &       $1.2\times10^{-2}$  \\
$\sqrt{10^{-4}}$   &	 	$-65.80$       &       $0.776$		    &        $1.5\times10^{-2}$  
	   \end{tabular}\centering
            }
\caption{Statistics of $\zeta$ for $W(\vp,\chi)=W_0\chi^ae^{-\lambda\vp^2}$ for different decay rates. All decay rates are in units of $\sqrt{W_0}\Mp$. We give values computed at the end of inflation ($t_e$) and at the completion of reheating (final) where $\zeta$ is conserved. \textit{Left Table}: Quadratic minimum ($a=2$); $\lambda=0.06$, $\vps=10^{-3}\Mp$ and $\chis=16.0\Mp$.  \textit{Right Table}: Quartic minimum ($a=4$); $\lambda=0.055$, $\vps=5\times10^{-4}\Mp$ and $\chis=23.0\Mp$.}
\label{tab:oneminstats}
    \end{center}
\end{table}
However, the expression for $\nz$, Eq.~(\ref{eq:index}), can be reduced to a function of solely $\Nvp$, by making use of the scaling relations Eqs.~(\ref{eq:derivs_scale}) and~(\ref{eq:derivs_scale2}). Using $g_*\equiv\Nchi\approx(2\ecstar)^{-1/2}$ and $\Nchichi\approx 1-(\etacc/2\ec)_*=1/2$ in Eq.~(\ref{eq:index}), with $(\dot{\vp}_I/H)_* \approx -(2\epsilon_I^*)^{1/2}$ and relations Eqs.~(\ref{eq:derivs_scale}) and~(\ref{eq:derivs_scale2}) we may write:
\be
\label{eq:index_2}
\nz-1 \approx -2\epsilon_* - \frac{2\Nvp^2}{\Nvp^2+g_*^2}\left[4\lambda-\frac{\sqrt{2\epstar}}{\vps}-\frac{4\lambda}{\Nvp}\frac{\sqrt{2\epstar}}{2\ecstar} + \frac{1}{2\Nvp^2}\right]\,.
\ee
By the completion of reheating, the last two terms in parenthesis are small, suppressed by factors of $\Nvp$ and may be neglected. With $\sqrt{2\epstar}/\vps=2\lambda$ we arrive at
\be
\label{eq:index_approximation_2}
\nz-1 \approx -2\epsilon^* - \frac{4\lambda\Nvp^2}{\Nvp^2+g_*^2} \geq  -2\epsilon^* -4\lambda\,.
\ee
This shows that there is a bound for the allowed range of $\nz$, and also explains the almost complete insensitivity of $\nz$ to $\Gchi$ when $\Nvp^2\gg g^2_*$, an example of which is given in Table~\ref{tab:oneminstats} for $\lambda=0.06$. By comparison, $\fnl$ is much more sensitive to $\Gchi$ for the same value of $\lambda$, since $\fnl(\Nvp)$ is not flat over the $\Nvp$ range of interest. This indicates that, for this particular inflationary model, $\nz$ is a more robust inflationary observable, and perhaps a better probe of the underlying inflationary model since it is insensitive to the physics of reheating. Whilst the two models (with quadratic or quartic minima) studied in this section can generate a large $\fnl$ that survives until the completion of reheating, they would be ruled out by observation since their spectral indices are far too low.\\

The qualitative arguments given in this section are respected as long as $\lambda$ is not too large. If $\vp\sim\mathcal{O}(\lambda^{-1/2})$, then the scaling relations Eqs.~(\ref{eq:derivs_scale}) and~(\ref{eq:derivs_scale2}) will break down and our arguments may no longer hold. The effect of reheating on the motion of $\vp$ is to prevent it from rolling any further down its potential: as reheating becomes more efficient, $\vp$ freezes out at smaller values. In this strong coupling regime, we expect the scaling relations to work well, but becoming a worse approximation if reheating proceeds very slowly.

\section{Two Minima}\label{sec:twomins}

Assisted inflation~\cite{liddle:1998} may be realised via a collection of string axions. In this scenario, known as N--flation~\cite{Dimopoulos:2005}, the many axion fields \textit{cooperatively} source inflation even if their potentials are individually too steep. The collective potential is comprised of a sum of $N_f$ uncoupled axions $\vp_i$:
\be
\label{eq:Nflation}
W(\vp)=\sum_{i=1}^{N_f}\Lambda_i^4\left[1 -{\rm cos}\left(\frac{2\pi}{f_i}\vp_i\right)\right]\,.
\ee
With only a single field present, this model is more commonly known as natural inflation~\cite{Adams1993Natural}. Each axion is fully described by its decay constant $f_i$ and its potential energy scale $\Lambda_i^4$. The standard arguments show that we should expect $f_i\sim10^{16}{\rm GeV}$. The mass of each field in vacuum satisfies $m_{\vp(i)}^2=4\pi^2\Lambda_i^4/f_i^2$. Due to the shift symmetry $\vp_i\rightarrow\vp_i+n f_i$, we can without loss of generality set the initial conditions $\vp_{*(i)}\in[0,\,f_i]$. To generate a large $\fnl$, we must have at least one axion close to the `hilltop' at $\vp\sim0.5$~\cite{Kim:2011je}. The evolution of the curvature perturbation in the post-inflationary epoch, including the reheating stage, for sum-separable multifield models was also studied by Choi et.al. \cite{Choi:2008et}. However, these authors restricted themselves to scalar fields with quadratic minima and similar masses which decay with similar rates; while in the following we consider a more general scenario and include the studies of non--gaussianity $\fnlf$ as well as the tensor--to--scalar ratio $r$.

\subsection{Quadratic minimum} \label{sec:quadtwomins}

We follow~\cite{elliston:2011}, supposing that the initial conditions are chosen so that only a single axion $\vp$ populates this hilltop region. This field sources the non--Gaussianity, whilst the remaining $N_f-1$ axions, which begin away from the hilltop, contribute only to the expansion rate. By expanding about the minimum of the remaining $N_f-1$ fields,  these axions may be replaced by a single effective field $\chi$ with a quadratic potential. With $f_i=f$ for all axions, the effective two--field potential then reads:
\be
\label{eq:effectiveNflation}
W(\vp,\chi)=W_0\left[\frac{1}{2}m^2\chi^2 + \Lambda^4\left(1 -{\rm cos}\left(\frac{2\pi}{f}\vp\right)\right)\right]\,.
\ee
In fact, replacing the collective potential with an effective two-field potential is well motivated, see for example \cite{Battefeld:2008bu}, where they showed that the energy density of the universe is dominated by fields with comparable masses even if one starts with thousands of fields, including the post-inflationary reheating stage. Reheating in models of N-flation also proceeds preferentially via a perturbative decay route as opposed to via parametric resonance and preheating \cite{Battefeld:2008bu,Battefeld:2009xw}.

From this point onwards we will refer to $\vp$ as the axion and to $\chi$ as the inflaton. By suitably choosing the axion/inflaton mass ratio in vacuum, various scenarios can be realised. For example, if the axion is sufficiently massive it may quickly decay to its minimum during inflation, where it becomes trapped without oscillating. In this case, adiabaticity is established long before reheating begins, and the decay of the inflaton into radiation does not affect the evolution of $\zeta$. We have confirmed this numerically. 

\begin{figure*}[t]
	\begin{tabular}{cc}
		\includegraphics[width=0.48\linewidth]{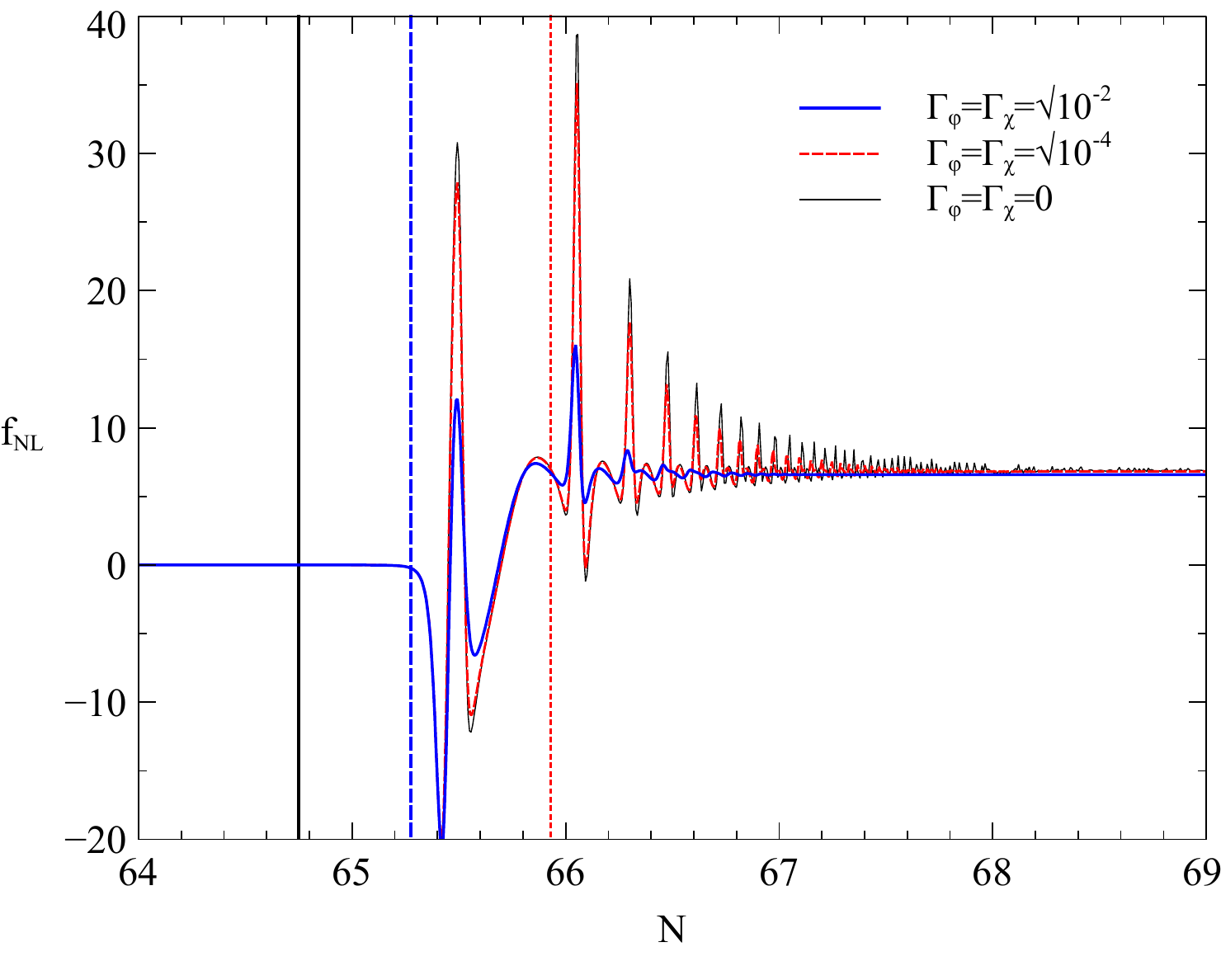}
		\includegraphics[width=0.48\linewidth]{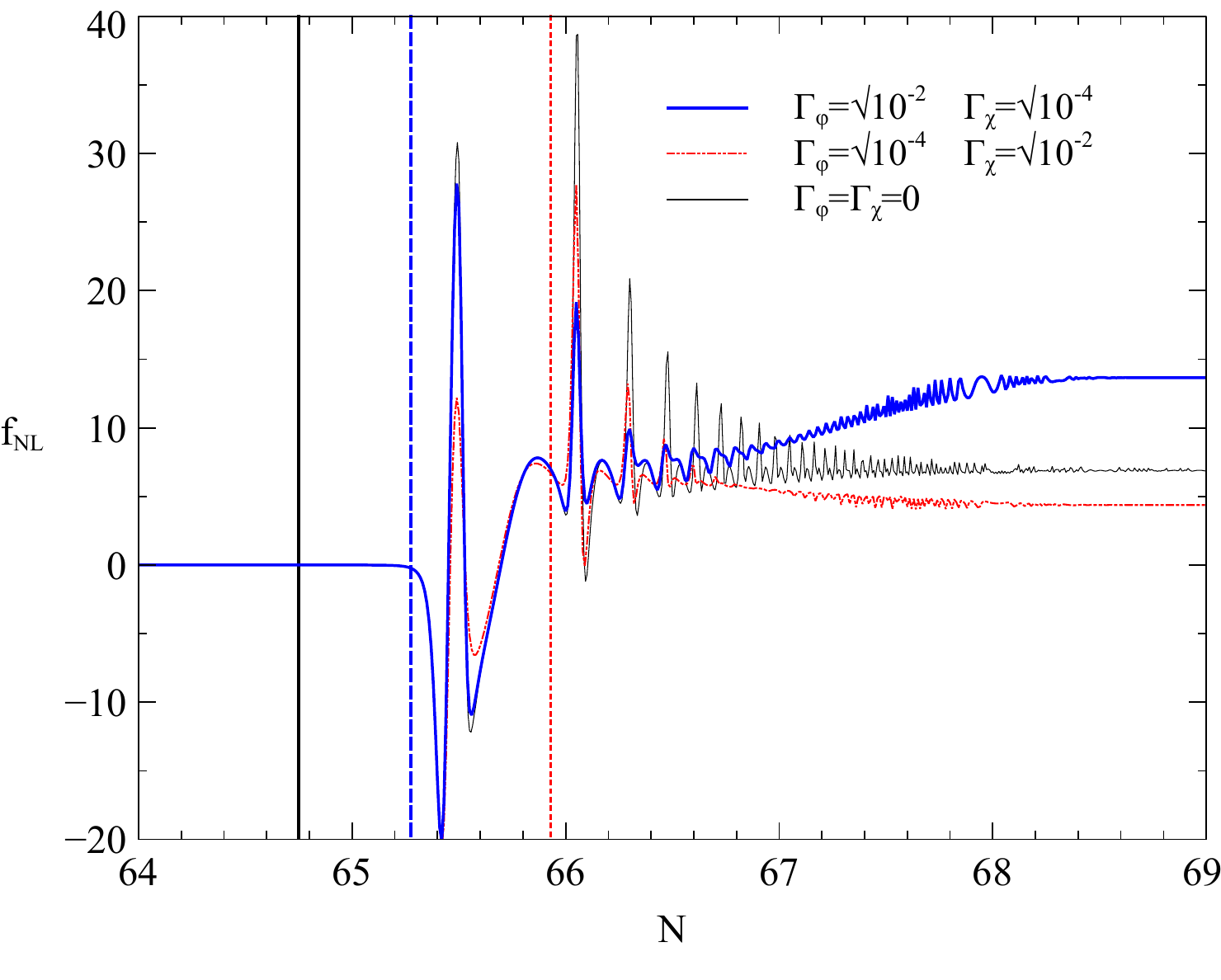}
	\end{tabular}	
\caption{$W(\vp,\chi)=W_0\left[\frac{1}{2}m^2\chi^2 + \Lambda^4\left(1 -{\rm cos}\left(\frac{2\pi}{f}\vp\right)\right)\right]$. The parameters used are: $\Lambda^4=m^2f^2/4\pi^2$, $\vps=(\frac{f}{2}-0.001)\Mp$, $\chi_*=16\Mp$, $f=m=1$. Both panels show the evolution of $\fnl$ during reheating. \textit{Left Panel}: Equal decay rates, $\Gchi=\Gvp\neq0$. For comparison we also show the $\Gchi=\Gvp=0$ limit (thin black line). \textit{Right Panel}: Unequal decay rates, $\Gchi\neq\Gvp\neq0$. For comparison we also show the $\Gchi=\Gvp=0$ limit (thin black line). In both panels, the solid vertical (black) line denotes the end of inflation, $N_{\rm e}$, the dashed vertical (blue) line denotes the start of $\chi$ reheating and the dotted vertical (red) line denotes the start of $\vp$ reheating.  The background Hubble rates at the $\chi$ and $\vp$ reheating surfaces are $H_r^\chi\approx\sqrt{5\times10^{-2}W_0}\Mp$ and $H_r^\vp\approx\sqrt{10^{-2}W_0}\Mp$ respectively.} 
\label{fig:axion_fNL_1}
\end{figure*}

It is also possible to realise dynamics where both fields minimise after inflation has ended, entering an oscillating phase such that perturbative reheating can be applied. For example, with $\Lambda^4=m^2f^2/4\pi^2$, $\vps=(\frac{f}{2}-0.001)\Mp$, $\chi_*=16\Mp$ and $f=m=1$, the inflaton minimises before the axion, but both fields minimise after inflation has ended. In this example both fields acquire the same mass in vacuum. Fig.~\ref{fig:axion_fNL_1} shows the evolution of $\fnl$ for different combinations of $\Gchi$ and $\Gvp$ for this parameter choice. Since both fields oscillate rapidly about their minima, both fields must be coupled to radiation. If one field is instead left uncoupled, its energy density will scale as matter since the minimum is quadratic, and will eventually come to dominate over radiation which redshifts away more quickly. 

Unlike the product separable case where the universe is reheated from only a single field, the $\vp$ field has left slow--roll by the time reheating starts. Hence, the non--linear dynamics during the oscillating phase is essential and we could not find any simple scaling relation between $\Nvpvp$, $\Nchivp$ and $\Nvp$. Yet we find that $\fnl$ is still dominated by the same term as in the case where adiabaticity is reached before inflation ends~\cite{elliston:2011}:
\be
 \label{eq:fnl_axion}
  \fnl \approx \frac{5}{6}\frac{\Nvpvp}{\Nvp^{2}}\,.
\ee

As can be seen from the left panel of Fig.~\ref{fig:axion_fNL_1}, $\fnlfinal$ is almost completely insensitive to reheating when $\Gchi\sim\Gvp$. However, as can be seen from the right panel, a mild hierarchy between $\Gchi$ and $\Gvp$ generates significant corrections to to $\fnlfinal$. This effect is \textit{not} due to the axion reheating hypersurface being distinctly separated from the inflaton surface (the vertical dotted (red) and dashed (blue) lines of Fig.~\ref{fig:axion_fNL_1} respectively) and we have confirmed this numerically. What is important however, is the axion/inflation mass ratio in vacuum. The model parameters which realise the dynamics seen in Fig.~\ref{fig:axion_fNL_1} give $m_\vp=m_\chi$ at the minimum. The differences induced in $\fnlfinal$ when a mild hierarchy exists between $\Gchi$ and $\Gvp$ is greatest when the masses are equal. As the masses are separated, keeping the ratio $\Gchi/\Gvp$ fixed, the sensitivity of $\fnlfinal$ to reheating decreases. This can be understood as follows: first consider the situation where the two fields have different masses, for instance, $m_\chi>m_\vp$. Assuming both fields reheat at roughly the same time, the more massive field $\chi$ will dominate the energy density of the universe and thus the dynamics of the universe during reheating. Evaluating on constant energy hypersurfaces, the initial horizon crossing dependence of the $\chi$ field dynamics is smaller compared to the case $m_\chi=m_\vp$, where the energy density of the universe is distributed evenly between the fields. As a result, we expect the number of $e$--folds of expansion $N$ and $\fnlfinal$ are less sensitive in the case $m_\chi\neq m_\vp$.

In fact, having the two fields decay at different rates is a form of modulated reheating, although it is different from the standard scenario~\cite{Kofman2003Probing,Dvali2004New,Suyama2008NonGaussianity}. In the standard modulated reheating scenario, inflation is driven by a single field, whose decay rate is modulated by a second, subdominant field that remains light and plays a negligible role during inflation. The fluctuations of the subdominate field induce fluctuations in the inflaton decay rate and thus generate curvature perturbation during reheating. In the two minima case here, note that the initial horizon crossing values of the fields $\vps,\chis$ determine how the energy density of the universe is distributed between the two scalar fields. Therefore, although the field decay rates are constant here, the rate of energy transfer from the scalar fields to the radiation fluid can be different for each inflationary trajectory in the bundle and thus can generate extra contributions to the curvature perturbation, provided there is a mild hierarchy in the decay rates. Therefore it is not surprising that $\fnl$ can acquire such a significant correction during reheating when the two decay rates are different. The two minima scenario is also similar in spirit to a model of two field inflation with equal masses followed by instant preheating, in which the two field have very different couplings to the preheat field~\cite{Byrnes2006Scaleinvariant}, for a related scenario see also~\cite{Kolb2005Curvature}. Note however that all of these instant preheating models are very tightly constrained even at the level of linear perturbations~\cite{Byrnes2009Constraints}.

\subsection{Quartic minimum} \label{sec:quartwomins}

We now repeat the same analysis, promoting the quadratic $\chi^2$ minimum of Eq.~(\ref{eq:effectiveNflation}) to a quartic minimum, $\chi^4$. This modification was also studied in~\cite{Elliston2012Large} where the model parameters were chosen such that $\zeta$ becomes conserved during slow--roll.  Again, we find a similar qualitative behaviour of $\fnlfinal$ as in the quadratic case: the asymptotic values of $\fnl$ are very insensitive to the decay rates of the scalar fields when they are equal, and slightly more sensitive if they are different. However, all observables are much less sensitive to decay rates here as compared to the quadratic minimum case. 

We summarize the values of the observables of $\zeta$ for the models studied in Section~\ref{sec:quadtwomins} and~\ref{sec:quartwomins} at the end of inflation and end of reheating in Table~\ref{tab:twominstats}.

\renewcommand*\arraystretch{1.2}
\begin{table}[h!]
\vspace{5pt}
    \begin{center}
        \subtable{
\begin{tabular}{c|c|c|c|c}
\multicolumn{5}{c}{$\chi^2$ minimum: $\fnl(t_e)\approx0$, } \\
\multicolumn{5}{c}{$n_s(t_e)=0.969$, $r(t_e)=0.124$} \vspace{2mm}  \\
\hline
\hline
$\Gvp$               &         $\Gchi$               &      $\fnlfinal$  &    $n_s^{\rm final}$   &  $r^{\rm final}$ \\
\hline
$0$   		   &              $0$   		&	 	$6.88$       &       $0.935$		    &        $4.6\times10^{-4}$  \\
$\sqrt{10^{-2}}$   &              $\sqrt{10^{-2}}$   &	 	$6.59$       &       $0.969$		    &        $4.3\times10^{-4}$  \\
$\sqrt{10^{-4}}$   &		$\sqrt{10^{-4}}$   &	 	$6.83$       &       $0.965$		    &        $4.6\times10^{-4}$  \\
$\sqrt{10^{-2}}$   &		$\sqrt{10^{-4}}$   &	 	$13.66$     &       $0.963$		    &        $1.0\times10^{-3}$  \\
$\sqrt{10^{-4}}$   &		$\sqrt{10^{-2}}$   &	 	$4.37$       &       $0.974$		    &        $2.7\times10^{-4}$  
                \end{tabular}\centering
            }
\hspace{10mm}
        \subtable{
\begin{tabular}{c|c|c|c|c}
\multicolumn{5}{c}{$\chi^4$ minimum: $\fnl(t_e)\approx0$, } \\
\multicolumn{5}{c}{$n_s(t_e)=0.951$, $r(t_e)=0.263$} \vspace{2mm}  \\
\hline
\hline
$\Gvp$               &         $\Gchi$               	&      $\fnlfinal$      &    $n_s^{\rm final}$   &  $r^{\rm final}$ \\
\hline
$0$   		   &              $0$   		&	 	$5.04$       &       $0.966$	    &        $2.9\times10^{-4}$  \\
$\sqrt{10^{-5}}$   &              $\sqrt{10^{-5}}$   &	 	$4.99$       &       $0.972$	    &        $3.0\times10^{-4}$  \\
$\sqrt{10^{-4}}$   &		$\sqrt{10^{-4}}$   &	 	$5.06$       &       $0.966$	    &        $3.0\times10^{-4}$  \\
$\sqrt{10^{-1}}$   &		$\sqrt{10^{-5}}$   &	 	$5.39$       &       $0.967$         &        $3.3\times10^{-4}$  \\
$\sqrt{10^{-2}}$   &		$\sqrt{10^{-4}}$   &	 	$5.28$       &       $0.967$         &        $3.2\times10^{-4}$ 
                \end{tabular}\centering
            }
\caption{Statistics of $\zeta$ for $W(\vp,\chi)=W_0\left[\frac{1}{2}m^2\chi^a + \Lambda^4\left(1 -{\rm cos}\left(\frac{2\pi}{f}\vp\right)\right)\right]$ for different decay rates. All decay rates are in units of $\sqrt{W_0}\Mp$. We give values computed at the end of inflation ($t_e$) and at the completion of reheating (final) where $\zeta$ is conserved. \textit{Left Table}: Quadratic minimum ($a=2$); $\Lambda^4=m^2f^2/4\pi^2$, $\vps=(\frac{f}{2}-0.001)\Mp$, $\chi_*=16\Mp$, $f=m=1$.  \textit{Right Table}: Quartic minimum ($a=4$); $\Lambda^4=m^2f^2/4\pi^2$, $\vps=(\frac{f}{2}-0.001)\Mp$, $\chi_*=22\Mp$, $f=m=1$. Notice the very large decrease in the tensor--to--scalar ratio from the end of inflation to its final value.}
\label{tab:twominstats}
    \end{center}
\end{table}

\section{Non-separable potential with one minimum}\label{sec:non_separable}

In previous sections, we have studied the evolution of $\fnl$ and its asymptotic value at the end of reheating, $\fnlfinal$, in examples where one or both fields reheat from a two--field \textit{separable} potential. In this section, we will repeat the same analysis, but this time for a non--separable potential.

As an example, we consider a modified version of the previously studied quartic exponential model, by adding an extra quadratic mass term 
\be
 \label{eq:quar_nsp}
  W(\chi,\vp) = W_{0}(\chi^{4}e^{-\lambda\vp^2/\Mp^2} + \kappa\chi^2) \,.
\ee
Before discussing reheating, it is useful to first study the inflationary regime. During inflation, the quadratic $\chi^2$ mass term has a negligible effect on the field dynamics when the $\chi$ field is of $O(1)$ in Planckian units, unless $\kappa\gg O(1)$ or $\lambda\vp^{2}\gg O(\Mp^2)$. Here in the following, we only consider the case $\kappa\sim O(1)$, for which we can approximate the field dynamics and $\fnl$ during inflation as the same as setting $\kappa=0$. Therefore, in the region of parameter space where $\kappa\leq O(1)$, $\fnl$ is expected to follow similar evolution as in the separable case studied in Section~\ref{sec:oneminquartic} during the slow-roll regime, with large deviations only coming in at late times towards the end of inflation.

The mechanism for generating large $\fnl$ is the same as discussed in \cite{elliston:2011}, which is well illustrated from the fact that there exists a scaling relation for the subdominate field $\Dn$ derivatives.

For the values $\kappa=1$, $\lambda=0.05$, $\vps=10^{-3}\Mp$, $\chis=22\Mp$, a large negative $\fnl$ is generated during inflation as the $\vp$ field rolls down the ridge and the bundle of trajectories diverge. The evolution is similar to the separable case where $\kappa=0$, with $\fnl\approx -44$ close to the end of slow-roll. Things are however a bit different after inflation even before reheating starts. For $\lambda=0.06$, the additional quadratic term becomes comparable to the quartic term slightly earlier than in the case of $\lambda=0.05$. In this case, we find $\fnl$ swaps sign shortly after the end of inflation. 
This unexpected behaviour, which we do not see in other cases, may be explained as follows: although the trajectories are still diverging in the $\vp$ direction in this case, the fact that the quadratic term becomes dominant suggests that the local potential geometries around each trajectory converge to the same quadratic shape, independent of $\vp$. This would have the same effect as the trajectories themselves converging in the separable case where $H$ is converging, thus giving momentarily large positive $\fnl$.

Shortly after inflation ends, when the $\chi$ field reaches sub--Planckian values, the $\chi^2$ term starts to dominate over the $\chi^4$ term. Therefore, we expect the additional $\chi^2$ term modifies the field dynamics during the reheating phase and possibly $\fnl$ as well. The additional $\chi^2$ term makes the potential less shallow around the minimum. This saves the $\chi$ field from being frozen to non-zero values, leaving unwanted residual potential energy in the case where $\Gchi$ is too large where the oscillations of the scalar fields are heavily damped. This happens if the potential around the minimum is too shallow, as in the model studied in Section~\ref{sec:oneminquartic}.
\begin{figure*}[t]
	\begin{tabular}{cc}
		\includegraphics[width=0.48\linewidth]{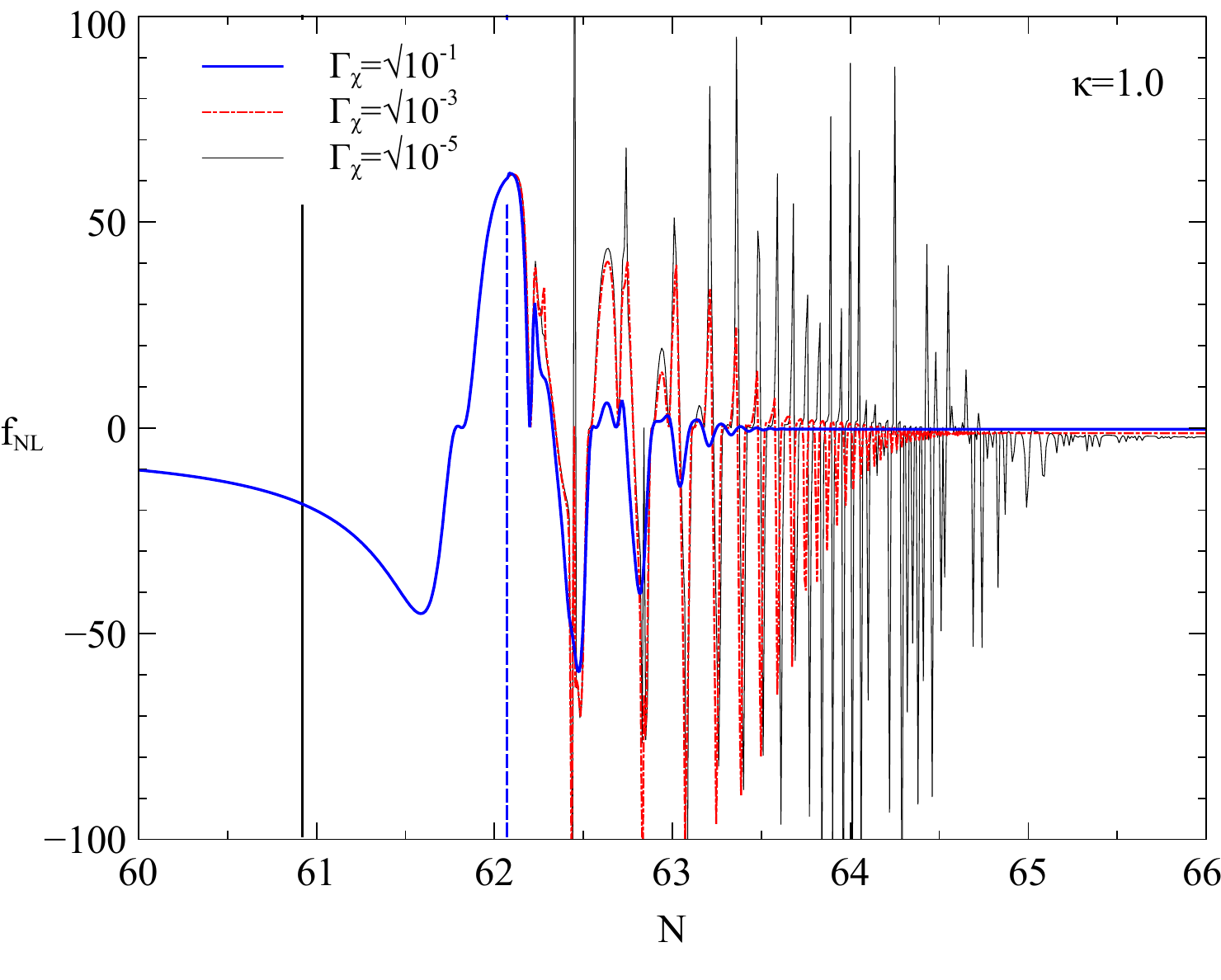}
		\includegraphics[width=0.48\linewidth]{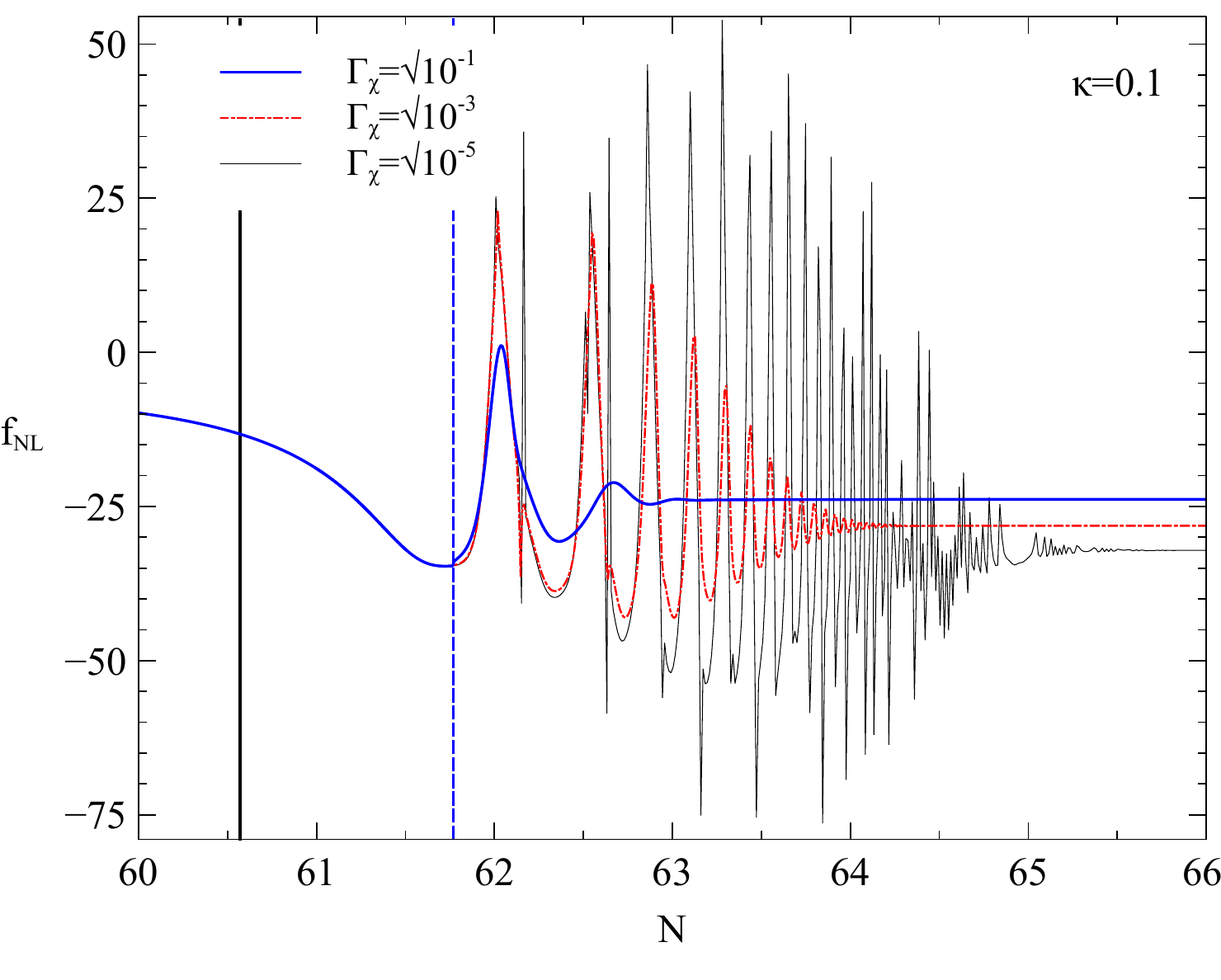}
	\end{tabular}
\caption{$W(\chi,\vp) = W_{0}(\chi^{4}e^{-\lambda\vp^2/\Mp^2} + \kappa\chi^2)$. We show $\fnl$ as a function of $N$ during reheating. The parameters used are: $\vps=10^{-3}\Mp$, $\chis=22\Mp$ and $\lambda=0.06$. In both panels, the solid vertical (black) line denotes the end of inflation, $N_{\rm e}$, and the dashed vertical (blue) line denotes the start of reheating, $N_{\rm r}$.  \textit{Left Panel}: $\kappa=1.0$. The Hubble rate at the start of reheating is $H_r\approx\sqrt{2\times10^{-1}W_0}\Mp$. \textit{Right Panel}: $\kappa=0.1$. The Hubble rate at the start of reheating is $H_r\approx\sqrt{10^{-1}W_0}\Mp$.}
\label{fig:fNL_nsp}
\end{figure*}
Similar to the separable case, as shown in Fig.~\ref{fig:fNL_nsp}, we found that $\fnl$ oscillates roughly in phase with $\chi^{2}$ during the early reheating stage, with a larger amplitude for smaller $\Gchi$. However, unlike the previous separable case in Section~\ref{sec:oneminquartic}, the $\Dn$ derivatives and $\fnl$ are now much less sensitive to $\Gchi$ and thus the reheating timescale. The sensitivity increases as $\kappa$ decreases as shown in Table~\ref{tab:nonSepStats}.
This means the effect of introducing additional quadratic mass term reduces the sensitivity of $\fnl$ to the reheating timescale. \footnote{Note that changing $\kappa$ also slightly changes the times that inflation ends and reheating starts. This would however have negligible effect on the dependence of the observables on $\Gchi$ in the parameter space of interest.}

\renewcommand*\arraystretch{1.2}
\begin{table}[h!]
\vspace{5pt}
    \begin{center}
        \subtable{
\begin{tabular}{c|c|c|c}
\multicolumn{4}{c}{Non--separable $\kappa=1.0$ $\fnl(t_e)=-18.71$, } \\
\multicolumn{4}{c}{$n_s(t_e)=0.748$, $r(t_e)=4.1\times10^{-3}$} \vspace{2mm}  \\
\hline
\hline
$\G_\chi$               &      $\fnlfinal$  &    $n_s^{\rm final}$   &  $r^{\rm final}$ \\
\hline
$\sqrt{10^{-5}}$   &	 	$-2.27$   &       $0.912$	    &        $2.0\times10^{-1}$  \\
$\sqrt{10^{-3}}$   &	 	$-1.28$   &       $0.896$	    &        $2.1\times10^{-1}$   \\
$\sqrt{10^{-1}}$   &	 	$-0.345$   &       $0.899$	    &        $2.1\times10^{-1}$  
                \end{tabular}\centering
            }
\hspace{10mm}
        \subtable{
\begin{tabular}{c|c|c|c}
\multicolumn{4}{c}{Non--separable $\kappa=0.1$ $\fnl(t_e)=-13.23$, } \\
\multicolumn{4}{c}{$n_s(t_e)=0.746$, $r(t_e)=2.0\times10^{-3}$} \vspace{2mm}  \\
\hline
\hline
$\Gchi$               &      $\fnlfinal$  &    $n_s^{\rm final}$   &  $r^{\rm final}$ \\
\hline
$\sqrt{10^{-5}}$   &	 	$-32.1$   &       $0.747$	    &        $1.5\times10^{-2}$  \\
$\sqrt{10^{-3}}$   &	 	$-28.1$   &       $0.752$	    &        $1.1\times10^{-2}$   \\
$\sqrt{10^{-1}}$   &	 	$-23.9$   &       $0.751$	    &        $7.8\times10^{-3}$  
	   \end{tabular}\centering
            }
\caption{Statistics of $\zeta$ for $W(\chi,\vp) = W_{0}(\chi^{4}e^{-\lambda\vp^2/\Mp^2} + \kappa\chi^2)$ for different decay rates. All decay rates are in units of $\sqrt{W_0}\Mp$. We give values computed at the end of inflation ($t_e$) and at the completion of reheating (final) where $\zeta$ is conserved. \textit{Left Table}: The parameters used are: $\vps=10^{-3}\Mp$, $\chis=22\Mp$ and $\lambda=0.06$ and $\kappa=1.0$.  \textit{Right Table}: The parameters used are: $\vps=10^{-3}\Mp$, $\chis=22\Mp$ and $\lambda=0.06$ and $\kappa=0.1$.}
\label{tab:nonSepStats}
    \end{center}
\end{table}
%
%

\section{Discussion and Conclusions}\label{sec:discConclu}

In this work we have studied the effects of perturbative reheating on the key inflationary observables $\fnl$, $\nz$ and $r$, for canonical two--field inflation models. We have considered two classes of potential: the `runaway' type which has a minimum in only one direction; and potentials which have a minimum in both directions. We have studied quadratic and quartic minima, finding that the dependence of the statistics of $\zeta$ on the decay rate of the field(s) is qualitatively the same. Perhaps most importantly, we have shown for both classes of models, that if a large non--Gaussian signal exists at the start of reheating, it will in general be non--zero at the completion of reheating.

For the single minimum models, adiabaticity is never established before reheating begins and so the bi--spectrum acquires substantial reheating--dependent corrections. As a consequence, the magnitude of any non--Gaussianity generated at the end of inflation does not necessarily remain the same at the end of reheating, meaning that $\fnl$ cannot be linked directly to the physics of the inflationary model. Whilst $\fnl$ is sensitive to reheating, we have also shown that there can exist certain regimes of model parameter space where the spectral index $\nz$ is almost completely insensitive to reheating. In such scenarios, $\nz$ may be considered a more robust inflationary statistic and a better probe of the underlying potential.

For two--minima models where both fields decay to reheat the universe, we have shown numerically that even if an adiabatic condition is \textit{approached} by the inflating/isocurvature fields converging in, and oscillating about, their global minima, the decay of these fields into radiation can promote further evolution of $\zeta$. If a mild hierarchy in decay rates between each field exists, $\fnl$ can be enhanced or suppressed relative to the same model where reheating is not accounted for.

\begin{figure*}[t]
	\begin{tabular}{cc}
		\includegraphics[width=0.48\linewidth]{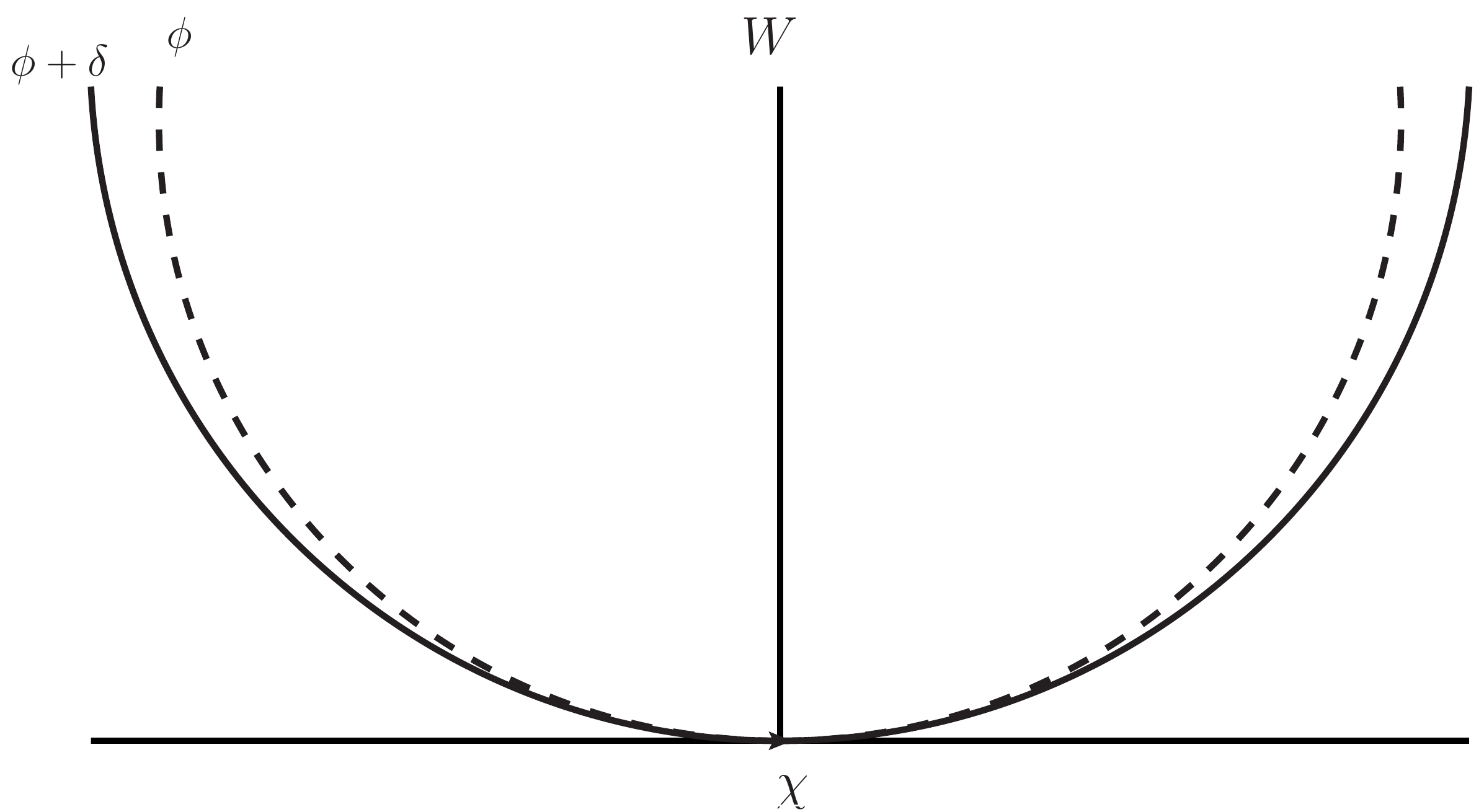}
		\includegraphics[width=0.48\linewidth]{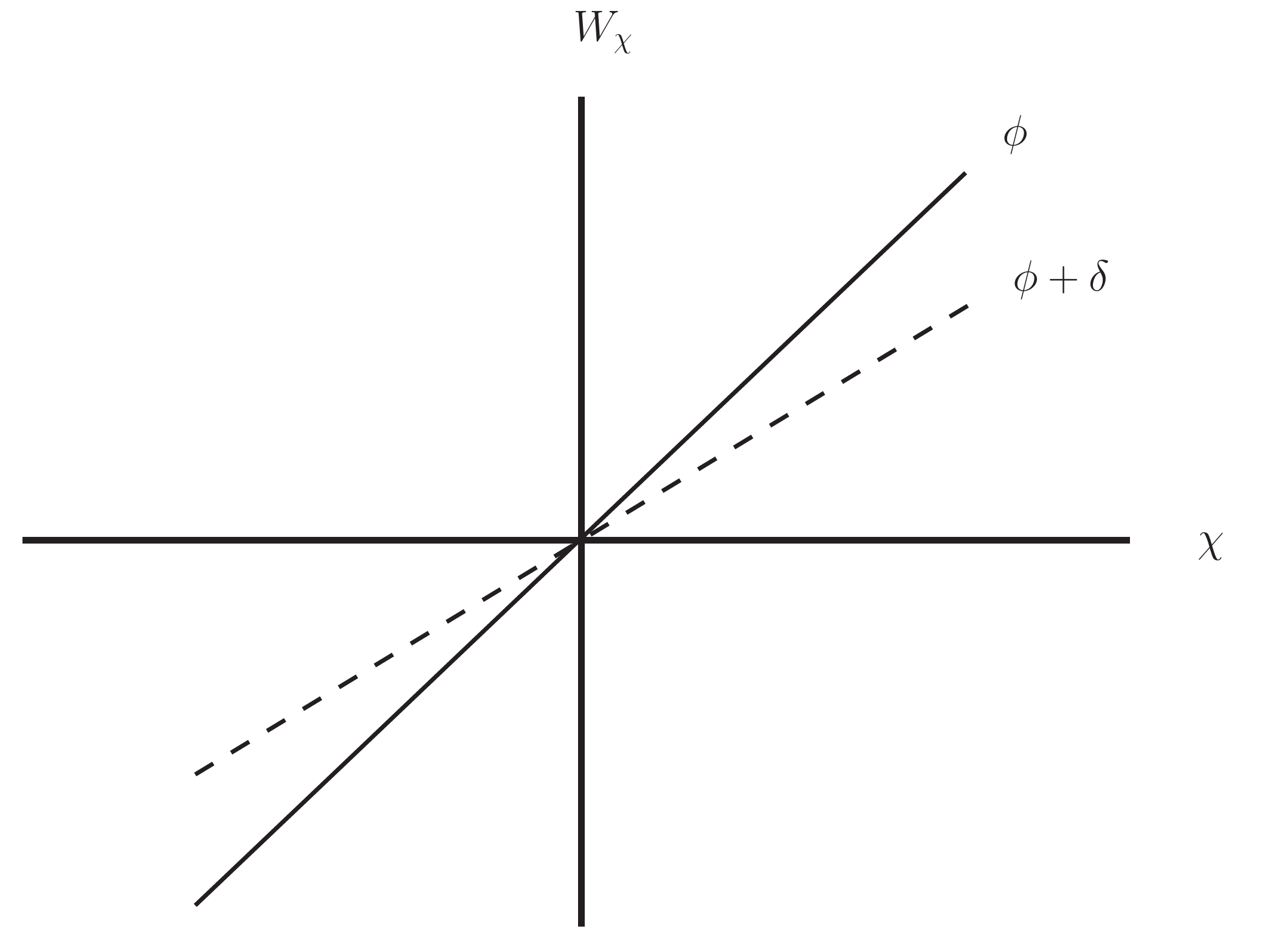}
	\end{tabular}	
\caption{An illustration of how the reheating minima depend on the subdominate field as in the case of quadratic times exponential potential. The perturbation in the $\phi$ direction is denoted by $\delta$. \textit{Left Panel}: Local geometries of the minima around the $\chi$ direction. \textit{Right Panel}: The potential gradient in the $\chi$ direction.} 
\label{fig:pot_minima}
\end{figure*}

One important difference between the single minimum models and the two minima model is that in the former case, the fields are coupled via the potential, whilst in the latter they are coupled only via gravity. Thus, for the single minimum models of Sections~\ref{sec:onemin} and \ref{sec:non_separable}, the local geometries of the $\chi$ minima are functions of the subdominate field $\vp$, and these geometries are different for different inflationary trajectories in the bundle. This is illustrated in Fig.~\ref{fig:pot_minima}. The shape of these `reheating minima' evolve in time as reheating proceeds, and will affect the dynamics of the oscillating $\chi$ field. In the two--minima model of Section~\ref{sec:twomins} however, where the potential is sum--separable and the fields are coupled only through gravity, the local geometries of the $\chi$ minima are always independent of $\vp$ and so this effect is not present. 
This effect is of course model dependent, as we have illustrated with the non--separable model Eq.~(\ref{eq:quar_nsp}).  When the interaction term is small and plays a negligible role during reheating, the $\chi$ field dynamics are independent of the dynamics of $\vp$. This explains why we found the sensitivity of the $\Dn$ derivatives to $\Gchi$ decreases as $\kappa$ increases. For larger $\kappa$, the $\vp$ dependence of the local geometries of the $\chi$ minima decreases and thus $\fnlfinal$ is very insensitive to $\Gchi$. 

In general, for both classes of potential (one and two minimum), the degree of sensitivity of the primordial observables to reheating seems to be a model (and parameter) dependent issue.  That said, for all the models studied in this work, whilst the magnitude of $\fnl$ may be heavily dependent on the decay rate of the field(s), its sign remains the same. Whether this conclusion encompasses more complicated models is unclear.  For example, couplings between the inflating and isocurvature fields may promote the global minima studied in this paper, to functions of the fields themselves (local minima) possibly leading to a highly non--trivial reheating surface which may generate large corrections to $\zeta$. Models with non--canonical kinetic terms, such as DBI inflation may also not respect this observation, since the fields no longer follow the gradient of the scalar potential.

Another point worthy of some discussion is that we have assumed reheating to take place entirely perturbatively. Whilst we consider this to be a sensible starting point for a first--time exploration of the sensitivity of $\fnl$, $\nz$ and $r$ to reheating, this is almost certainly a gross simplification: the initial stages of particle production (preheating) is a violently explosive non--perturbative effect. It is expected that such a rapid preheat stage in the regime of broad resonance may have long--lasting effects on the subsequent evolution of the universe. For example, it may lead to specific non--thermal phase transitions in the early universe,~\cite{Kofman1996NonThermal,Rajantie2000Phase}, topological defect production and promote novel mechanisms for baryogenesis~\cite{Anderson1996Preheating,Trodden2004Baryogenesis}. Preheating has also been shown to generate significant levels of non--Gaussianity~\cite{Chambers2008Lattice,Chambers2008NonGaussianity,Kofman2003Probing,Dvali2004New,Suyama2008NonGaussianity,Byrnes2009Constraints}. If a more sophisticated description of (p)reheating were employed, including the rich spectrum of perturbative and non--perturbative QFT effects, it is tempting to speculate that the statistics of $\zeta$ might be more radically altered in models of multi--field inflation. 

Non--Gaussianity has evolved into a very active and topical field, in which observations have improved greatly over the last decade, through both studies of the CMB and large scale structure. At the same time, on a theoretical and phenomenological level, a plethora of different mechanisms have been suggested which are capable of generating an observable $\fnl$. Currently, the tightest constraints on local type $\fnl$ come from the WMAP satellite, which constrains the amplitude of the non--Gaussian part of $\zeta$ to be less than about one thousandth of the amplitude of the Gaussian perturbation. Planck is expected to tighten this constraint considerably.  A detection of $\fnl$ at this level would rule out the simplest canonical, single field inflation models, where it has recently been explicitly shown that preheating has a negligible effect on the scalar bi--spectrum~\cite{Hazra2012Scalar}. However, we will most likely be left with many other viable scenarios, including multi--field models, which when suitably tuned, can match the observations. As we have demonstrated in this paper, accounting for the dynamics of reheating muddies the waters further. Unless adiabaticity has been achieved before the onset of reheating, it seems unlikely that we can use explicit values of $\fnl$ to discriminate between different multi--field models unless we have a complete understanding of the reheating process.\footnote{This applies to local--types of non--gaussianity only. For other shapes such as equilateral type, the post--inflationary evolution will not change the inflationary predictions as the contributions come from interactions at or before horizon--crossing, see for examples~\cite{Fergusson:2008ra,Battefeld:2011ut,Ribeiro:2011ax}} In this sense, our work lends support to that of~\cite{Barnaby2011Phenomenology}, in that it also represents a challenge to the conventional lore that non--Gaussianity is a `smoking gun' signature of non--standard inflationary dynamics: such signatures may be significantly altered by the subsequent reheating phase. 

More optimistically, non--Gaussianity is not only about one single number. The trispectrum (the four--point function) depends on two non--linearity parameters $\tau_{\rm NL}$ and $g_{\rm NL}$, and if the current observations $|\fnl|\sim40$ (which are not statistically significant) turn out to be true, then $\tau_{\rm NL}$ should be large enough for Planck to detect. It might be that, like the spectral index for the single minimum models studied in this paper, the trispectrum is less sensitive to the physics of reheating.  Furthermore, if $\fnl$ is detected, it may also be possible to constrain or even detect a scale dependence: $\fnl$ is often assumed to be constant, but this is only true for certain simple models. For example $\fnl$ is strongly scale dependent in the two--field hybrid inflation model~\cite{Byrnes2009Large}. This opens up the question of whether reheating may leave some observable signature in the running of $\fnl$, which may be used as a complimentary probe of the inflationary theory and the reheating mechanism itself. Indeed, whilst it has been shown that $\fnl$ is insensitive to preheating in canonical single field models (as well as being too small to be observed) it is strongly scale dependent~\cite{Hazra2012Scalar}. 

When discussing the sensitivity of the primordial observables to reheating, it is also important to keep in mind the degree of fine tuning which is required for the inflationary model itself to be consistent with observations. For typical models this amount of fine tuning is large, especially if one wishes to generate an observable $\fnl$. Accounting for the subsequent dynamics of reheating introduces a further source of fine tuning, however what is apparent from this work, is that this is secondary compared to the degree of inflationary fine tuning. For the single minimum models studied in this paper for example, changing $\lambda$ or $\vps$ by one part in $10^2$ may completely remove any non--Gaussian signal, whilst shifting the reheating decay rate by two orders of magnitude changes $\fnl$ by $\mathcal{O}(2)$ units.

In conclusion, whilst non--Gaussianity is in principle a powerful probe that may be used to distinguish between the many models of inflation, we must be careful in our interpretation of any observational constraints that place bounds on the statistics of $\zeta$. We have shown that the dynamics of perturbative reheating can have a non--negligible impact on these statistics for canonical two--field inflation models. As such, without a UV complete theory of inflation \textit{and} reheating, it seems hard to infer the properties of the underlying inflationary potential from observational bounds on $\fnl$ and related quantities alone.

\acknowledgements

The authors would like to thank David Seery, David Lyth, Thorsten Battefeld, Filippo Vernizzi, Paul Saffin and David Mulryne for useful discussions. GL and ERMT are supported by the University of Nottingham. EJC acknowledges the STFC, Royal Society and Leverhulme Trust for financial support.

\bibliographystyle{apsrev}
\bibliography{ppxet}

\end{document}